\documentstyle[12pt]{article}
\input{epsfig.sty}
\newcommand{\beq}{\begin{eqnarray}}
\newcommand{\eeq}{\end{eqnarray}}
\begin{document}
\title{Review of QCD, Quark-Gluon Plasma, Heavy Quark Hybrids, and Heavy Quark
State production in p-p and A-A collisions}
\author{Leonard S. Kisslinger$^{1}$\\
Department of Physics, Carnegie Mellon University, Pittsburgh, PA 15213\\
Debasish Das$^{2,3}$\\
High Energy Nuclear and Particle Physics Division,\\
Saha Institute of Nuclear Physics,1/AF, Bidhan Nagar, Kolkata 700064, INDIA.}
\date{}
\maketitle
\vspace{-1cm}

\noindent
1) kissling$@$andrew.cmu.edu \hspace{1cm} 2)dev.deba$@$gmail.com; 
3) debasish.das@saha.ac.in

\begin{abstract}
  This is a review of the Quantum Chromodynamics Cosmological Phase Transitions,
the Quark-Gluon Plasma, the production of heavy quark states via p-p collisions
and RHIC (Relativistic Heavy Ion Collisions) using the mixed hybrid theory for 
the $\Psi(2S)$ and $\Upsilon(3S)$ states; and the possible detection of the 
Quark-Gluon Plasma via heavy quark production using RHIC. Recent 
research on fragmentation for the production of D mesons is reviewed, as is
future theoretical and experimental research on the Collins and 
Sivers fragmentation functions for pions produced in polarized p-p collisions.
\end{abstract}


\noindent
Charm/Bottom Quarks,mixed hybrid theory

\vspace{2mm}
\noindent
PACS Indices:12.38.Aw,13.60.Le,14.40.Lb,14.40Nd

\section{Outline of  QCD Review, QCDPT, Detection of Quark-Gluon
 Plasma}

\hspace{1.4cm}{\bf QCD Theory of the Strong Interaction}
\vspace{2 mm}

\hspace{8mm}{\bf The QCD Phase Transition (QCDPT)}
\vspace{2 mm}

\hspace{8mm}{\bf Heavy Quark Mixed Hybrid States}
\vspace{2 mm}

\hspace{8mm}{\bf Proton-Proton Collisions and Production of Heavy Quark States}
\vspace{2 mm}

\hspace{8mm}{\bf RHIC and Production of Heavy Quark States}
\vspace{2 mm}

\hspace{8mm}{\bf Production of Charmonium and Bottomonium States via
Fragmentation}
\vspace{2 mm}

\hspace{8mm}{\bf  Sivers and Collins Asymmetries With a Polarized Proton
Target}
 \vspace{2 mm}

\hspace{8mm}{\bf Brief Overview}

\section{Brief Review of Quantum Chromodynamics (QCD) }

  In the theory of strong interactions  quarks, fermions, interact via 
coupling to gluons, vector (quantum spin 1) bosons,
the quanta of the strong interaction fields, color replaces the electric
charge in QED, which is why it is called Quantum Chromodynamics or QCD.
See Refs\cite{gw1973},\cite{w1973},\cite{fgl73}, and Cheng-Li's book on
gauge theories\cite{cl1984}.

 The QCD Lagrangian is
\beq
\label{QCD-L}
      \mathcal{L}_{QCD}&=& -\frac{1}{2}tr[G_{\mu \nu}G^{\mu \nu}]
+\sum_{k} \bar{q}_k(i \gamma^\mu(\partial_\mu-i g A_\mu)-m_k)q_k \nonumber \\
          G_{\mu \nu}&=& \partial_\mu A_\nu-\partial_\nu A_\mu
-ig[A_\mu A_\nu-A_\nu A_\mu] \\
         A_\mu&=& \sum_{1}^{8} A_\mu^a \lambda^a/2 \nonumber \; , 
\eeq 
where $q_k$ is a quark field with flavor $k$ and $A_\mu^a$ is the strong
interaction field, called the gluon field, with the quanta called qluons,
$\gamma^\mu$ are the Dirac matrices, $a$ is color, and $g$ is the strong
interaction coupling constant.  The quark flavors are 
$q_k:u,d,s,c,b,t$=up, down, strange, charm, bottom, and top quarks; and 
$m_k$ are the quark masses. The quarks
which we shall call heavy quarks are charm (c) and bottom (b) quarks. Although
quark masses are not well defined, as one cannot make a beam of particles 
with color, the heavy quark masses are  $m_c\simeq$ 1.5 GeV and $m_b \simeq$ 
5.0 GeV.

  The $\lambda^a$ are the SU(3) color matrices, with
\beq
\label{lambda}
     \lambda^a \lambda^b-\lambda^b \lambda^a =i 2\sum_{c=1}^8 f^{abc} 
\lambda_c \; ,
\eeq
with $f^{abc}$ the SU(3) structure constants. The nonvanishing $f^{abc}$ are:
\beq
\label{fabc}
        f^{123}&=& 1, {\rm \;\;\;} f^{458}=f^{678}=\sqrt{3}/2, \nonumber \\
        f^{147}&=& f^{165}=f^{246}=f^{257}=f^{345}=f^{376}=1/2 \; .
\eeq

  The most important states with which we consider are mesons, which in
the standard model consist of a quark and antiquark. For example, the state
$|J/\psi(1S)>\propto |c\bar{c}(1S)>$, a charm-anticharm state, with a mass
of about 3.1 GeV, approximately the mass of two charm quarks. Other
states very important for this review are the Upsilon states  
$|\Upsilon(mS)>$, which in the standard model are $|b\bar{b}(mS)>$, with 
m=1,2,3.

  The quarks have a strong interaction by coupling to gluons. They also
have an electric charge and experience an electromagnetic force. This is
a much more familiar force than QCD. The quantum field theory, QED, is similar 
to QCD, with a Lagrangian
\beq
\label{QED-L}
    \mathcal{L}_{QED}&=& i\bar{\psi}(i\gamma^\mu(\partial_\mu-i e A^{EM}_\mu)-m)
\psi
\; ,
\eeq
where $\psi$ is a quantum field with electric charge $e$ and $A^{EM}_\mu$ is
the electromagnetic quantum field. The quantum of $A^{EM}_\mu$ is the photon,
which is much more familiar than the gluon
\clearpage

  As shown in the figures 1 and 2, the electromagnetic interaction with
$e^2 \simeq 1/137$ is weak enough so the lowest order Feynman diagram
illustrated in Fig.1 gives almost the entire electric force, while
$g^2 \simeq 100 \times e^2$ is so large that Feynman diagrams are not
useful. Nonperturbative theories, such as QCD sum rules discussed below,
must be used.

 The lowest order Feynman diagrams for two quarks interacting via the 
electromagnetic interaction and strong interaction are illustrated in 
Fig.1 and Fig. 2 below
\vspace{3cm}

\begin{figure}[ht]
\begin{center}
\epsfig{file=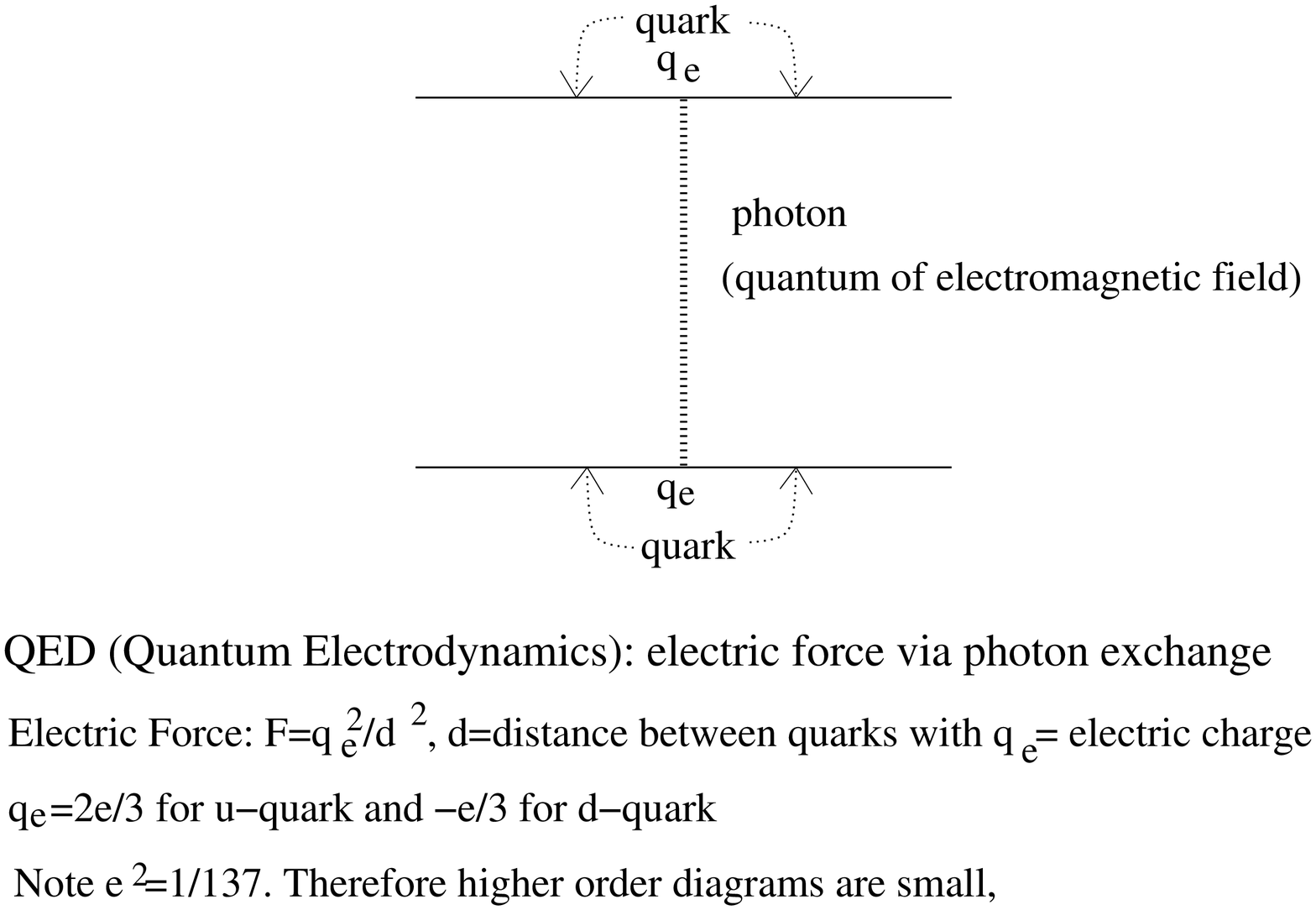,height=4cm,width=12cm}
\end{center}
\caption{Two quarks interacting via the exchange of a photon}
\end{figure}
\vspace{3cm}

\begin{figure}[ht]
\begin{center}
\epsfig{file=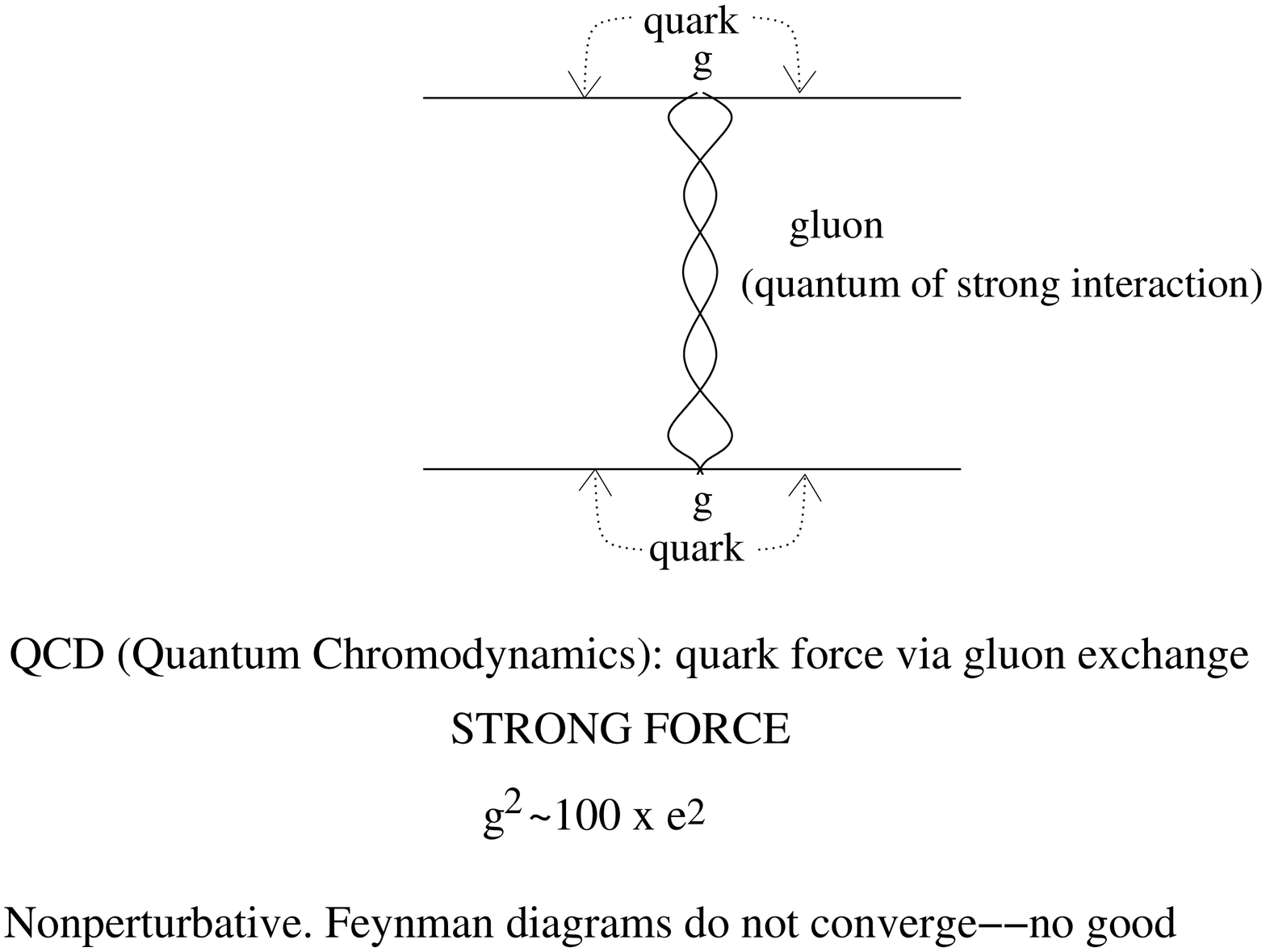,height=4cm,width=12cm}
\end{center}
\caption{Two quarks interacting via the exchange of a gluon}
\end{figure}
\newpage
\section{QCD Phase Transition}

A phase transition is the transformation of a system with a well defined
temperature from one phase of matter to another. The two basic types of
phase transitions are classical, when one phase transforms to another,
and quantum, when a state transforms to a different state. 

 The three most common classical phases are solid, liquid, and gaseous; and 
under special conditions there is a plasma phase. For early universe phase 
transitions the plasma phase is very important as the matter in the universe 
before the QCD phase transition was the Quark-Gluon Plasma, the main topic in 
this review. These classical phase transitions are illustrated in Figure 3.
\vspace{-4cm}

\begin{figure}[ht]
\begin{center}
\epsfig{file=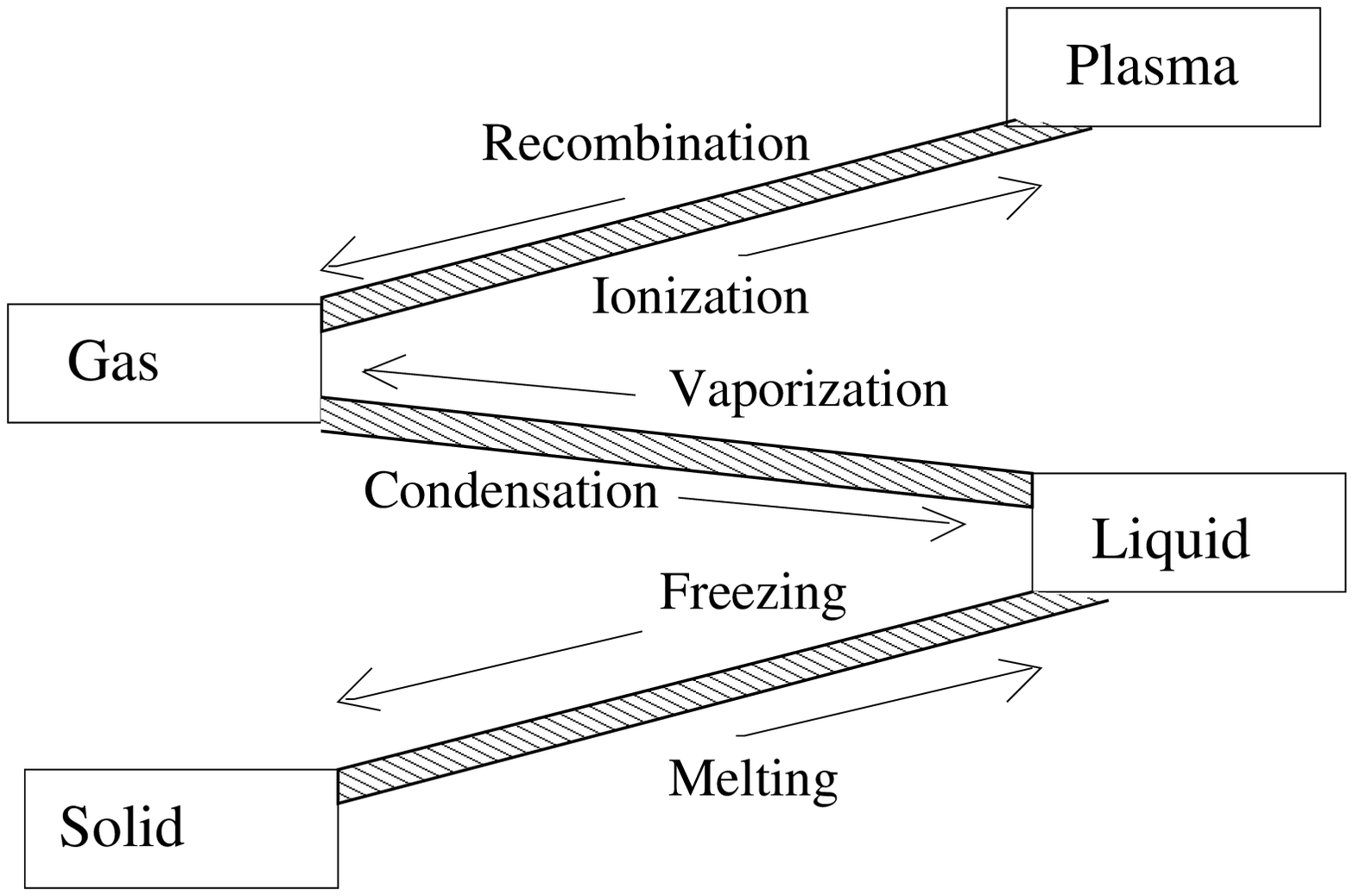,height=12cm,width=12cm}
\end{center}
\caption{Classical phases and phase transitions}
\end{figure}
\vspace{1cm}

  In the figure above, the ``Recombination'' transition is from a plasma
to a gas.  For the QCD Cosmological phases from a transition, discussed later 
in this section, as the Temperature of the universe dropped the matter
went from a Quark-Gluon Plasma to our present universe of protons and neutrons,
which is a gas (neither solid nor liquid), and later formed atomic nuclei
during the first 10-100s (see Fig. 4 on Evolution of the Universe below). 

   As we discuss in later sections, a major project of high energy nuclear
physics is to form the Quark-Gluon Plasma via the collisions of atomic
nuclei such as Copper (CU), lead (Pb), and gold (Au), and to detect it
by studying the production of heavy quark states.

   Next we briefly describe the evolution of the universe.
\clearpage

The universe has evolved for about 13.7 billion years. It has gone from a
very dense universe with very high temperature to our present universe,
with a number of important cosmoligacal events, as illlustrated in Fig 4.
\vspace{8cm}
\begin{figure}[ht]
\begin{center}
\epsfig{file=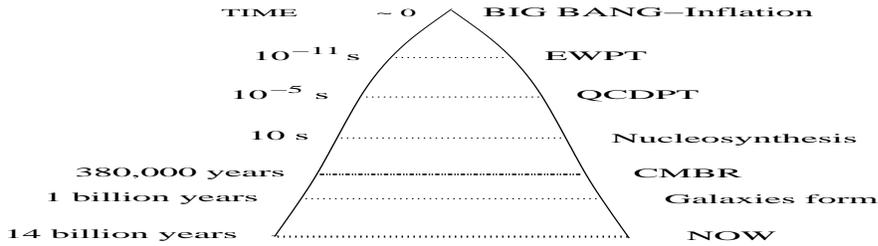,height=4cm,width=12cm}
\end{center}
\caption{Evolution of the Universe}
\end{figure}

  Inflation and Dark energy, which we do not discuss, occured at about
$10^{-34}$ seconds. The Electroweak Phase Transition (EWPT) occured at
at a time about $10^{-11}$ seconds after the big bang when the temperature
(a form of energy, so we use energy units) was T $\simeq$ 125 GeV, the
mass of the Higgs particle (discussed below). During the EWPT it is beleived 
that all particles except the photon got their mass. The QCD Phase Transition 
(QCDPT), the main topic in this review, occured at $t\simeq 10^{-5}$ s, with 
$T\simeq$ 150 MeV.

   The main event that we discuss in this review is the QCDPT.
Over three decades ago QCD and possible phase transitions at high T and 
density were discussed\cite{shur80}.  Inflation,
the EWPT, CMBR (Cosmological Microwave Background Radiation (from which
the amount of Standard and Dark Mass and Dark Energy have been measured)
 and events that occured after the QCDPT are discussed in detail in a
recently published book\cite{lskbook}.
\clearpage

\subsection{Classical Phase Transitions and Latent Heat}

  During a first order phase transition, with a critical temperature $T_c$,
as one adds heat the temperature stays at $T=T_c$ until all the matter has
changes to the new phase. The heat energy that is added is called latent
heat. This is illustrated in Fig 5.

  In contrast to a first order phase transition, a crossover transition
is a transition form one phase to another over a renge of temperatures,
with no critical temperature or latent heat.
\vspace{6cm}

\begin{figure}[ht]
\begin{center}
\epsfig{file=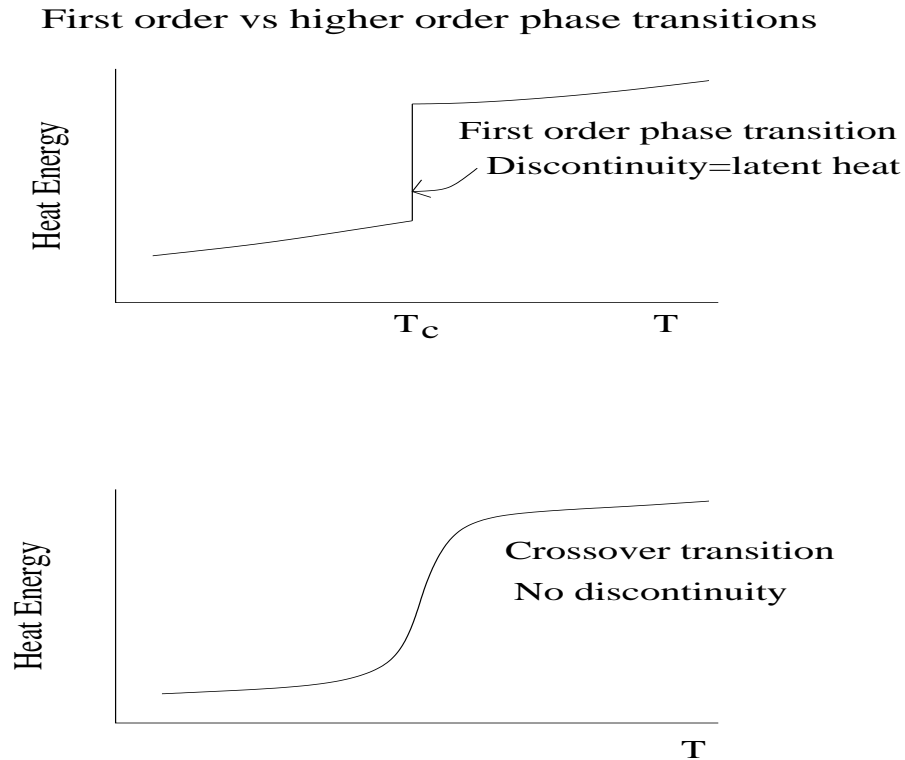,height=10cm,width=12cm}
\end{center}
\caption{First order and crossover phase transitions}
\end{figure}
\clearpage

 For application to cosmology
we are mainly interested in first order phase transitions. These phase
transitions occur at a critical temperature, $T_c$, and the temperature stays
the same until all matter in the system changes to the new phase. For example
if one heats water (a liquid) at standard atmospheric pressure it starts
to boil, with bubbles of steam (a gas), and the temperature stays at 
100 $C^o$. The heat energy that turns water to steam is called LATENT HEAT.
This illustrated in  Figure 5 above.

  A familiar example of first order phase transitions is ice, a solid, melting
to form water, a liquid; and water boiling to form steam, a gas. Figure 6
shows these two first order phase trasitiions for one gallon of 
water.  Note that the latent heat for ice-water and water-steam (water vapor) 
is given in calories. Recognizing that heat is a form of energy, in our
discussion of cosmological phase transitions we use units of energy for
ther latent heat.
\vspace{6cm}
\begin{figure}[ht]
\begin{center}
\epsfig{file=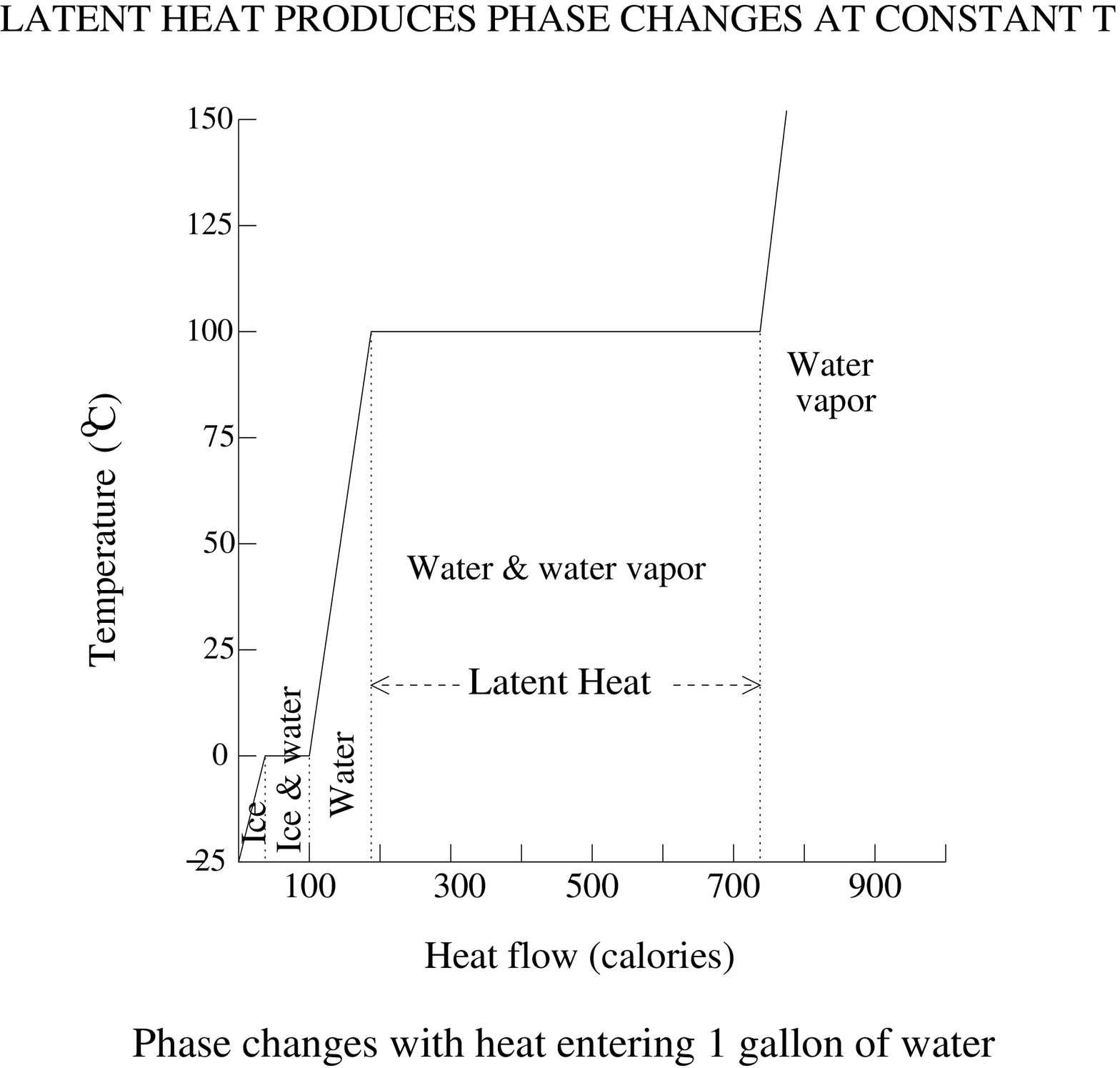,height=6cm,width=12cm}
\end{center}
\caption{Latent heat for ice to water and water to steam}
\end{figure}
\newpage
\subsection{Quantum Phase Transitions}

\subsubsection{Brief Review of Quantum Theory}

 In quantum theory one does not deal with physical matter, but with states and
operators. A quantum phase transition is the transition from one state to a 
different state. For the study of Cosmological Phase Transitions a
state is the state of the universe at a particular time and temperature.

   We now review some basic aspects of quantum mechanics needed for quantum
phase transitions.  A quantum state represents the system, and a quantum 
operator operates on a state. For instance, a system is in state [1] and
there is an operator $A$.
\beq
        |[1]> &\equiv& {\rm state[1]} \nonumber \\
           A &\equiv& {\rm operator\;A} \; .
\eeq

  An operator operating on a quantum state produces another quantum state.
For example, operator $A$ operates on state [1]
\beq
      A|[1]> &=& |[2]> \; ,
\eeq
where state [2]=$|[2]>$ is a quantum state. 

  State [2] might also be the same as state [1], with $|[1]>=|[2]>\equiv|A>$,
except for normalization,
\beq
      A|A> &=&  a|A> \; ,
\eeq
where $a$ is called the eigenvalue of the operstor $A$ in state $|A>$. It
is the exact value of $A$. If a state is not an eigenstate of an operator,
the operator does not have an exact value.

   In general, if a system is in a quantum state, the value of an operator
is given by the expectation valus. For example, consider state $|[1]>$ and
operator $A$.
\beq
        <[1]|&\equiv& {\rm adjoint\;of\;state[1]} \nonumber \\
        <[1]|A|[1]> &\equiv& {\rm expectation\;value\;of\;A} \; .
\eeq

     For example, classically an electron has momentum $\vec{p}$. In quantum
theory the system is in a state $|e,\vec{p}>$. The momentum operator when
operating on $|e,\vec{p}>$:
\beq
     \vec{p}_{op} |e,\vec{p}> &=& \vec{p}  |e,\vec{p}>\; ,
\eeq
since  $|e,\vec{p}>$  is  an eigenstate of the operator $\vec{p}$.

In quantum theory both position $\vec{r}$ and momentum $\vec{p}$ are
operators, with $p_x=(\not h/i)(d/dx)$, where $\not h=h/(2\pi)$ with $h$
Planks constant. Since $p_x x\neq x p_x$, a state cannot be an eigenstate of
both position and momentum. If the uncertainties in x,$p_x$ is $\Delta x,
\Delta p_x$ satisfy
\beq
         \Delta p_x  \Delta x &\geq& \not h/2 \; ,
\eeq
which is the Heisenberg Uncertainty Principle.
\newpage

\subsubsection{Cosmological Phase Transitions}

  Calling $|0,T>$ the state of the universe at time t when it has temperature
$T$, an operator $A$ has the expectation value $<0,T|A|0,T>$, as discussed
above. If there is a cosmological first order phase transition, then there
is a critical temperature $T_c$ and
\beq
\label{DeltaA}
        <0,T|A|0,T>_{T<T_c}- <0,T|A|0,T>_{T>T_C}&=& \Delta A \; ,
\eeq
with $ \Delta A$ the latent heat of the cosmological phase transitions.
The two very important cosmological phase transitions are the Electroweak
and QCD.

  The Electroweak Phase Transition (EWPT) took place at a time 
$t\simeq 10^{-11}$ seconds after the Big Bang, when the critical temperature was
$kT_c\simeq 125 GeV$. The operator A in Eq(\ref{DeltaA} is the Higgs field 
$\Phi$. $<0,T|\Phi|0,T>_{T>T_C}=0$, so the latent heat for the EWPT is
\beq
\label{higgs}
        <0,T|\Phi|0,T>_{T<T_c}&\propto& 125 GeV \simeq M_H \; ,
\eeq
with the Higgs particle recently detected at the LHC at CERN, with the
mass $M_H \simeq$ 125 GeV measured by the CMS\cite{CMS14} and 
ATLAS\cite{ATLAS14} collaborations. 
During the EWPT all standard model particles got their masses. With an
additional scalar field in the standard model, usually called the Stop,
the EWPT is first order, with baryogenesis (the creation of more quarks than 
antiquarks).

  The QCD Phase Transition (QCDPT), which is the main topic in this review,
took place at $t\simeq 10^{-5}$ seconds after the Big Bang, when the critical 
temperature was $kT^{QCDPT}_c\simeq 150 MeV$. It is a first order phase transition
and bubbles of our present universe with protons, neutrons, etc (hadrons)
nucleated within the universe with a dense plasma of quarks and gluons,
the Quark-Gluon Plasma (QGP) that existed when the temperature of the universe
was greater than $T^{QCDPT}_c$. This is illustrated in Fig. 7 We shall
discuss the possible detection of the QGP via heavy ion collisions.
\vspace{3.5cm}

\begin{figure}[ht]
\begin{center}
\epsfig{file=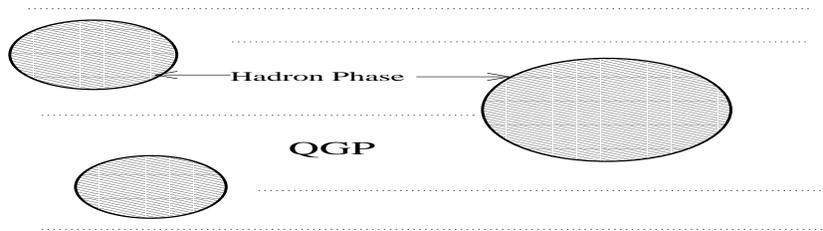,height=3cm,width=12cm}
\end{center}
\caption{Hadron phase forming within the QGP during the QCDPT}
\end{figure} 
\newpage

\subsection{The QCDPT and Quark Condensate}

  As reviewed above, the QCD fermion fields and particles are quarks. The
Latent Heat for the QCD Phase Transition (QCDPT) is the Quark Condensate,
which we now define.
\beq
\label{quark}
           q(x) &=& {\rm \;quark\;field} \nonumber \\
        \bar{q}(x) &=& {\rm \;antiquark\;field} \nonumber \\
              |0,T > &=& {\rm\;vacuum\;state\;temperature=T} \nonumber \\
      <0,T|\bar{q}(x) q(x)|0,T> &=& {\rm \;quark\;condensate} \nonumber \\
              &=& {\rm \;vacuum\;expectation\;value\;of\;\bar{q}(x) q(x)}
\nonumber
\eeq

\beq
  <0,T|\bar{q}(x) q(x)|0,T>&=& 0 {\rm \;in\;quark\;gluon\;plasma\;phase\;
T>T^{QCDPT}_c}
\nonumber \\
           &\simeq& -(.23\;GeV)^3 {\rm \;in\;hadron\;phase\;T<T^{QCDPT}_c} 
\nonumber
\eeq

  The QCDPT is first order, with a discontinuity on the quark condensate at
critical temperature. In Fig. 8 the results of a recent lattice 
gauge calculation for $<\bar{q}q>$, the quark condensate, are shown.
\vspace{2cm}

\begin{figure}[ht]
\begin{center}
\epsfig{file=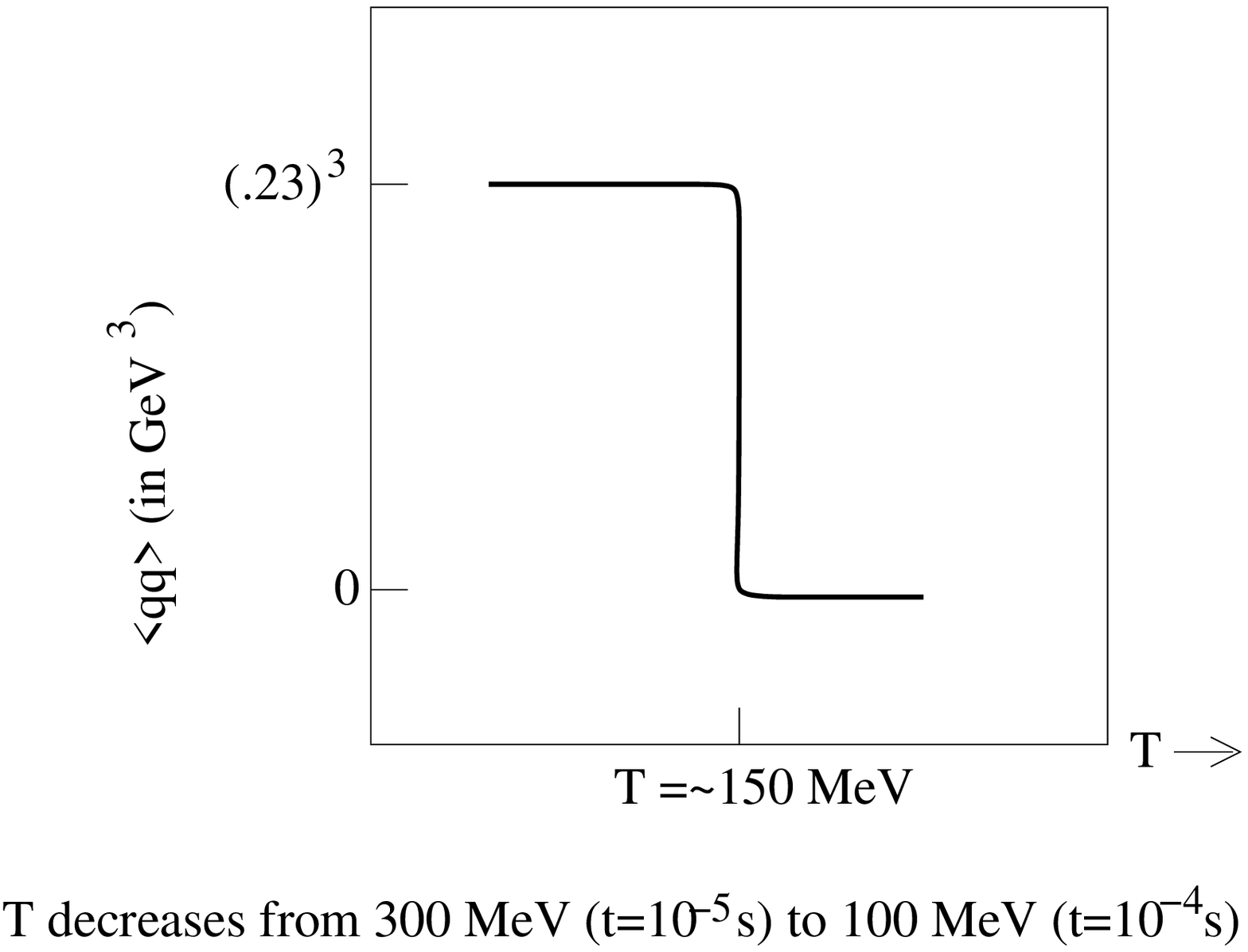,height=6cm,width=12cm}
\end{center}
\caption{The quark condensate as a function of T=temperature}
\end{figure} 

As one can see from the figure, the quark condensate $<\bar{q}q>$ goes
from 0 to $(.23)^3 GeV^3$ at the critical temperature of about 150 MeV,
and is therefore a first order phase transition. 
 
 Although we do not discuss Dark Energy in this review, note that
Dark Energy is cosmological vacuum energy, as is the quark condensate. It
has been shown that Dark Energy at the present time might have been created
during the QCDPT via the quark condensate\cite{zmk12}.

\newpage

\section{Review of mixed hybrid heavy quark meson states}

   The Charmonium and Upsilon (nS) states which are important for this review
 are shown in Fig. 9.
\vspace{4.5cm}

\begin{figure}[ht]
\begin{center}
\epsfig{file=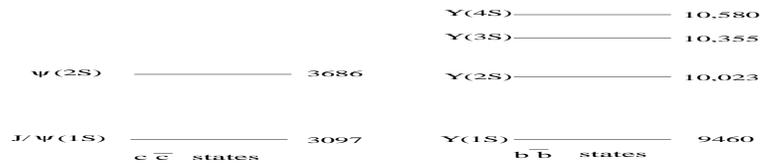,height=2cm,width=10cm}
\caption{Lowest energy Charmonium and Upsilon states}
%\label{1}
\end{center}
\end{figure}
\subsection{Heavy quark meson decay puzzles}

  Note that the standard model of the $\psi'(2S)$ and $\Upsilon(3S)$ as
$c \bar{c}$ and $b \bar{b}$ mesons is not consistent with the  following
puzzles:
   
 1) The ratio of branching rarios for $c\bar{c}$ decays into hadrons (h)
given by the ratios (the wave functions at the origin canceling)
\beq
  R&=&\frac{B(\Psi'(c\bar{c})\rightarrow h)}{B(J/\Psi(c\bar{c})\rightarrow h)}
\;=\;\frac{B(\Psi'(c\bar{c})\rightarrow e^+e^-)}{B(J/\Psi(c\bar{c})\rightarrow 
e^+e^-)}\simeq 0.12 \nonumber \; ,
\eeq
the famous 12\% RULE. 

  The $\rho-\pi$ puzzle: The $\Psi'(2S)$ to $J/\Psi$ ratios for $\rho-\pi$ 
and other h decays are more than an order of magnitude too small. 
Many theorists have tried and failed to explain this puzzle.
\clearpage

2) The Sigma Decays of Upsilon States puzzle: The $\sigma$ is a broad 600 MeV 
$\pi-\pi$ resonance.
\vspace{5mm}

$\Upsilon(2S) \rightarrow \Upsilon(1S) + 2\pi$ large branching ratio. 
No $\sigma$ 
\vspace{5mm}

$\Upsilon(3S) \rightarrow \Upsilon(1S) + 2\pi$ large branching ratio to 
$\sigma$ 
\vspace{5mm}

We call this the Vogel $\Delta n=2$ Rule\cite{vogel}.  
Neither of these puzzles can be solved using standard QCD models. They were
solved using the mixed heavy hybrid theory.
\subsection{Hybrid, mixed heavy quark hybrid mesons, and the puzzles}

   The method of QCD Sum Rules\cite{sz79} was used to study the heavy quark
Charmonium and Upsilon states, and show that two of them are mixed hybrid
meson states\cite{lsk09}, which we now review.
\subsubsection{Method of QCD Sum Rules}

   The starting point of the method of QCD sum rules\cite{sz79} for finding 
the mass of a state A is the correlator,
 
\beq
\label{2PiA(x)}
       \Pi^A(x) &=&  \langle | T[J_A(x) J_A(0)]|\rangle \; ,
\eeq
with $| \rangle$ the vacuum state and
the current $J_A(x)$ creating the states with quantum numbers A:
\beq
\label{JAx}
     J_A(x)| \rangle &=& c_A |A \rangle + \sum_n c_n |n; A  \rangle  \; ,
\eeq
where $ |A \rangle$ is the lowest energy state with quantum numbers A,
and the states $|n; A  \rangle$ are higher energy states with the A quantum
numbers, which we refer to as the continuum.     

   The QCD sum rule is obtained by evaluating $\Pi^A$ in two ways. First, 
after a Fourier transform to momentum space, a dispersion relation gives the 
left-hand side (lhs) of the sum rule:
\beq
\label{PiA}
 \Pi(q)^A_{\rm{lhs}} &=&  \frac{\rm{Im}\Pi^A(M_A)}
{\pi(M_A^2-q^2)}+\int_{s_o}^\infty ds \frac{\rm{Im}\Pi^A(s)}
{\pi(s-q^2)}
\eeq
where $M_A$ is the mass of the state $A$ (assuming zero width)
and $s_o$ is the start of the continuum--a parameter to be determined.
The imaginary part of $\Pi^A(s)$, with the term for the state we are
seeking shown as a pole (corresponding to a $\delta(s-M_A^2)$ term in 
$\rm{Im}\Pi$), and the higher-lying states produced by $J_A$ known as the 
continuum
Next $ \Pi^A(q)$ is evaluated by an operator product expansion
(O.P.E.), giving the right-hand side (rhs) of the sum rule
\beq
\label{PiArhs}
  \Pi(q)_{\rm{rhs}}^A &=& \sum_k c_k(q) \langle 0|{\cal O}_k|0\rangle
 \; ,
\eeq
where $c_k(q)$ are the Wilson coefficients and $\langle 0|{\cal O}_k|0\rangle$
are gauge invariant operators constructed from quark and gluon fields,
with increasing $k$ corresponding to increasing dimension of ${\cal O}_k$.

  After a Borel transform, ${\mathcal B}$, in which the q variable is 
replaced by the Borel mass, $M_B$ (see Ref\cite{sz79}),
the final QCD sum rule, ${\mathcal B} \Pi_A(q)(LHS) = 
{\mathcal B} \Pi_A(q)(RHS)$, has the form
\beq
\label{PiBorel}
    && \frac{1}{\pi} e^{-M_A^2/M_B^2}
+ {\cal B} \int_{s_o}^\infty \frac{Im[\Pi_A(s)]}{\pi(s-q^2)} ds \nonumber \\
     &=& {\cal B} \sum_k c_k^A(q) <0|{\cal O}_k|0> \; .
\eeq

 This sum rule and tricks are used to find $M_A$, which should vary little
with $M_B$. A gap between $M_A^2$ and $s_o$ is needed for accuracy. If the 
gap is too large, the solution is unphysical.

\subsection{Mixed charmonium-Hybrid charmonium States}

 Recognizing that there is strong mixing between a heavy quark meson and
a hybrid heavy quark meson with the same quantum numbers (defined below),
the following mixed vector ($J^{PC}=1^{--}$) charmonium, hybrid charmonium 
current was used in QCD Sum Rules 
\beq
\label{11}
        J^\mu &=& b J_H^\mu + \sqrt{1-b^2} J_{HH}^\mu 
\eeq
with
\beq
\label{12}
          J_H^\mu &=& \bar{q}_c^a \gamma^\mu q_c^a \nonumber \\
          J^\mu_{HH} &=&  \bar{\Psi}\Gamma_\nu G^{\mu\nu} \Psi \; ,
\eeq 
where $\Psi$ is the heavy quark field, $\Gamma_\nu = C \gamma_\nu$,
$\gamma_\nu$ is the usual Dirac matrix, C is the charge conjugation operator,
and the gluon color field is
\beq
\label{Gmunu}
         G^{\mu\nu}&=& \sum_{a=1}^8 \frac{\lambda_a}{2} G_a^{\mu\nu}
\; ,
\eeq
with $\lambda_a$ the SU(3) generator ($Tr[\lambda_a \lambda_b]
= 2 \delta_{ab}$), discussed above.

Therefore the correlator for the mixed state:
\beq
\label{13} 
   \Pi_{H-HH}^{\mu\nu}(x) &=& <0|T[J^\mu(x) J^\nu(0)]|0>
\eeq
is 
\beq
\label{14}
   \Pi_{H-HH}^{\mu\nu}(x) &=& b^2  \Pi_{H}^{\mu\nu}(x) + (1-b^2)
\Pi_{HH}^{\mu\nu}(x) \nonumber \\
  && +2b\sqrt{1-b^2}\Pi_{HHH}^{\mu\nu}(x) \\
       \Pi_{H}^{\mu\nu}(x)&=& <0|T[J_H^\mu(x) J_H^\nu(0)]|0> \nonumber \\
   \Pi_{HH}^{\mu\nu}(x)&=& <0|T[J_{HH}^\mu(x) J_{HH}^\nu(0)]|0> \nonumber \\
   \Pi_{HHH}^{\mu\nu}(x)&=& <0|T[J_H^\mu(x) J_{HH}^\nu(0)]|0> \nonumber \; ,
\eeq
where $\Pi_{H}^{\mu\nu}(x)$ is the correlator for the standard $c\bar{c}$
charm meson, $\Pi_{HH}^{\mu\nu}(x)$ is the correlator for a hybrid charm
meson, with a valence gluon, and $\Pi_{HHH}^{\mu\nu}(x)$ is the correlator
for a charm meson-hybrid charm meson.

   It was necessary to carry out many QCD sum rule calculations to
determine the value of the parameter $b$, which gives the relative
probability of a normal to a hybrid meson.
\clearpage

The leading diagrams for the meson and meson-hybrid meson diagrams are
shown in Fig. 10.
\vspace{4.3cm}

\begin{figure}[ht]
\begin{center}
\epsfig{file=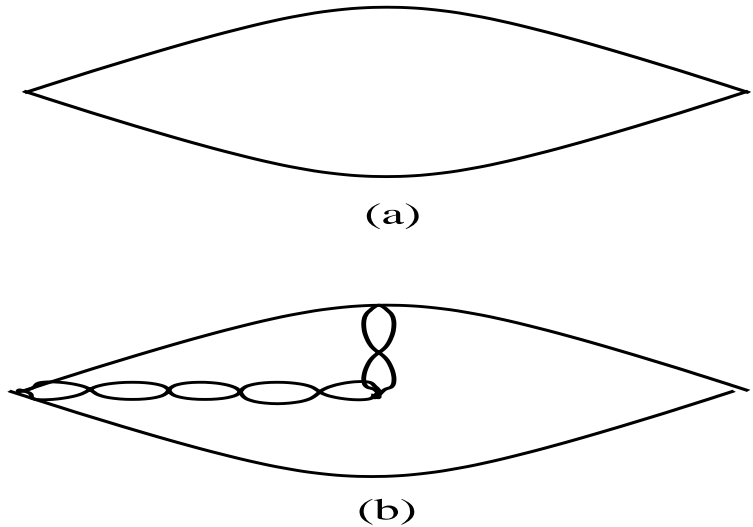,height=7cm,width=10cm}
\caption{(a) lowest order diagram for a heavy meson (b) lowest order diagram 
for a meson-hybrid meson}
%\label{}
\end{center} 
\end{figure}

  After a Fourier transform to find the correlator in momentum space,
$\Pi_{H-HH}^{\mu\nu}(p)$, the standard procedure for QCD sum rules was 
carried out.
 
  Finally the Borel transform of $\Pi_{H-HH}^{\mu\nu}(p)$ was found, from
which the square of the mixed meson-hybrid 
meson mass as function of the Borel mass, $M^2_{H-HH}$ was found. 
The result is $M^2_{C-HC} \simeq 3.69$ GeV=energy of the $\Psi'(2S)$ state.
A similar QCD sum rule calculation bottom heavy quarks found that the
mixed upsilon-hybrid upsilon mass is $M^2_{\Upsilon-H\Upsilon} \simeq 10.4$ GeV=
energy of the $\Upsilon(3S)$ state. 

From this we conclude that the
$\Psi'(2S)$ and $\Upsilon(3S)$ states are mixed meson-hybrid meson states.
This is very important for the study heavy quark state production via 
proton-proton collisions and RHIC for the detection of the Quark-Gluon Plasma,
since a hybrid mesons have a valence gluons, as does the QGP.
\newpage

For the mixed Charmonium-hybrid charmonium mass, $M^2_{C-HC}$, the result 
of the QCD sum rule analysis is shown in Fig. 11 for $b^2=0.5$.
\vspace{-6cm}

\begin{figure}[ht]
\begin{center}
\epsfig{file=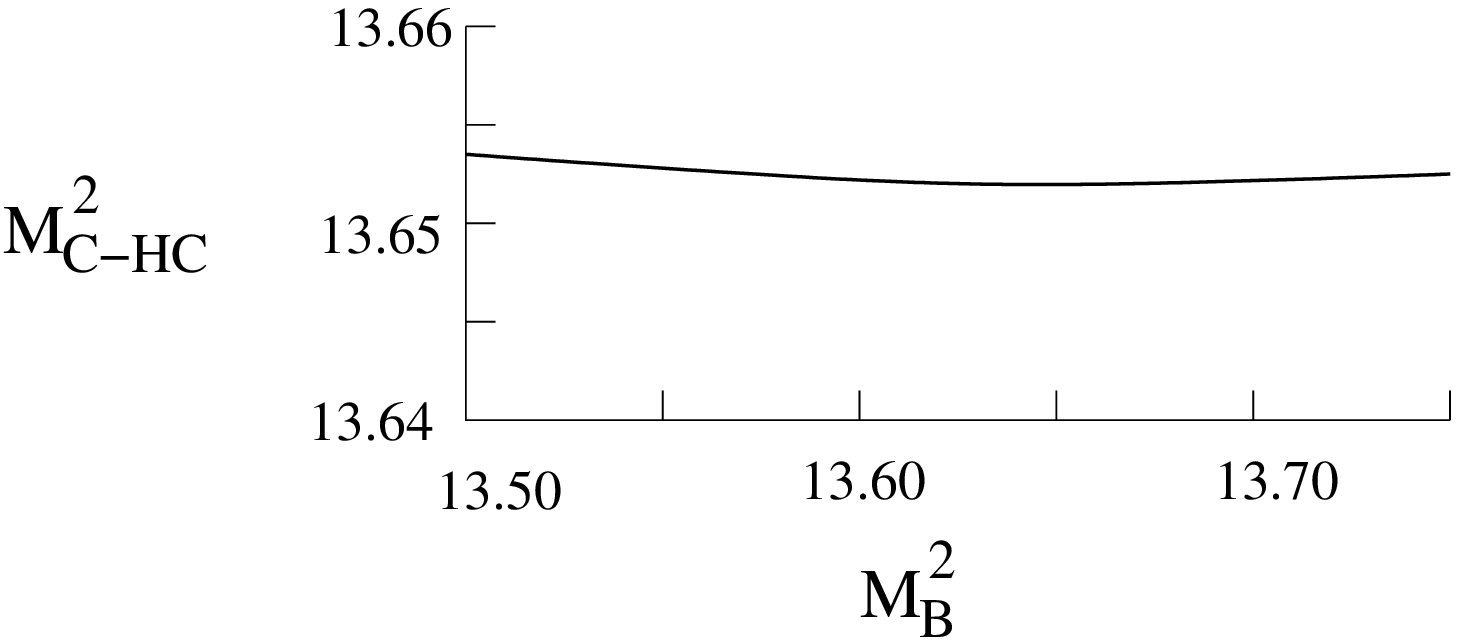,height=12cm,width=10cm}
\caption{Mixed Charmonium-hybrid charmonium mass $\simeq$ 3.69 GeV}
%\label{}
\end{center} 
\end{figure}

  From this figure one sees that the minimum in $M^2_{C-HC}(M_B^2)$ corresponds
to the  $\Psi'(2S)$ state being 50\% normal and 50\% hybrid. The
analysis for upsilon states was similar, with the $\Upsilon(3S)$ being 50\% 
normal and 50\% hybrid.

\section{Heavy Quark State Production In p-p Collisions}

There has been a great deal of interest in the production and polarization
of heavy quark states in proton-proton collisions. In additition to the
puzzles discussed above, the $J/\Psi,\Psi'$ production anomaly\cite{cdf97}, 
in which the charmonium production rate was larger than predicted for $J/\Psi$,
and much larger for $\Psi'$ than theoretical predictions in proton-proton (p-p)
collisions has motivated p-p heavy quark state production experiment. In 
addition to being an important study of QCD, these experiments also could 
provide the basis for testing the production of Quark-Gluon Plasma (QGP) via 
a Relativistic Heavy Ion Collider (RHIC).

At the proton-proton (p-p) energies of the Fermilab, BNL-RHIC, or the Large
Hadron Collider (LHC) the color octet dominates the color singlet model,
which we now review.

\subsection{Color Octet vs Color Singlet Heavy Quark State Production}

The color octet model was shown to dominate the color singlet 
model\cite{cl96,bc96,fl96}. We now discuss the Cho/Leibovich 
study\cite{cl96,cl296} which compared color octet to color singlet production. 
For the color singlet production they used the standard results of
Ref\cite{br73} and others with $\alpha_s=g^2/(4\pi)$, where $g$ is the strong 
coupling constant
$M=2M_Q$ and $q^0=\vec{q}^2/M$, with $\vec{q}$ the colliding particles momentum:
\beq
\label{singlet}
        \sigma(gg \rightarrow Q\bar{Q}[^1S^{(1)}_0])&=& \frac{\alpha^2_s M}
{384 \pi^2 q^0 s} \delta(1-M^2/s) \; ,
\eeq
\clearpage
The two color octet diagrams are shown in Fig. 12, with (a) 
representing quark-antiquark $\rightarrow$ gluon $\rightarrow$ color 8 
quark-antiquark state $\Psi_Q$; and (b) reprenting gluon-gluon $\rightarrow
\Psi_Q$.

\begin{figure}[ht]
\begin{center}
\epsfig{file=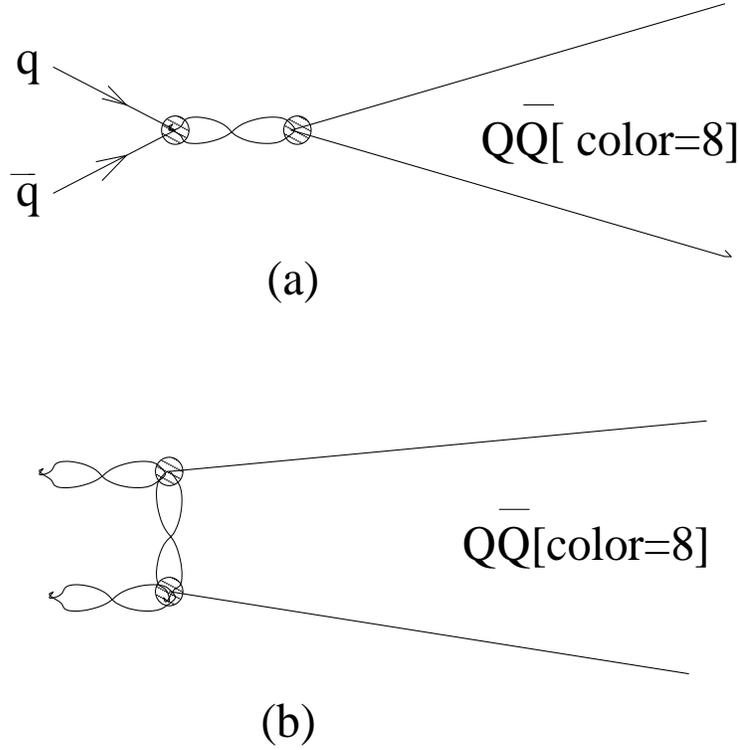,height=10cm,width=10cm}
\caption{Color octet diagrams for (a) $q\bar{q} \rightarrow \Psi_Q$(8) and (b)
$gg \rightarrow \Psi_Q$(8)}
%\label{}
\end{center} 
\end{figure}

The results for the $p-\bar{p} \rightarrow J/\Psi$ theotetical transverse 
momentum differential cross section for the singlet and octet theories 
and CDF data\cite{cl296} are shown in Fig. 13.  Solid curve  is 
color octet and dashed curve is color singlet production.

From this figure and references given above one sees that the color octet
theory dominates. As we shall see when discussing the theory of production
cross sections, there are a number of parameters that must be determined,
and the diagrams shown in the figure above are not simple Feynman diagrams 
from which one derives the matrix elements needed to predict the cross sections.

This rather complicated theory which we discuss in the next subsection
 is used for p-p production of heavy quark states,
which we discuss in the next subsection. It is also used in RHIC, AA production
of heavy quark states, as is iscussed in the following section.

\newpage

   Transverse momentum diffenrential cross section for $p-\bar{p} \rightarrow 
J/\Psi$:

\vspace{-2cm}

\begin{figure}[ht]
\begin{center}
\epsfig{file=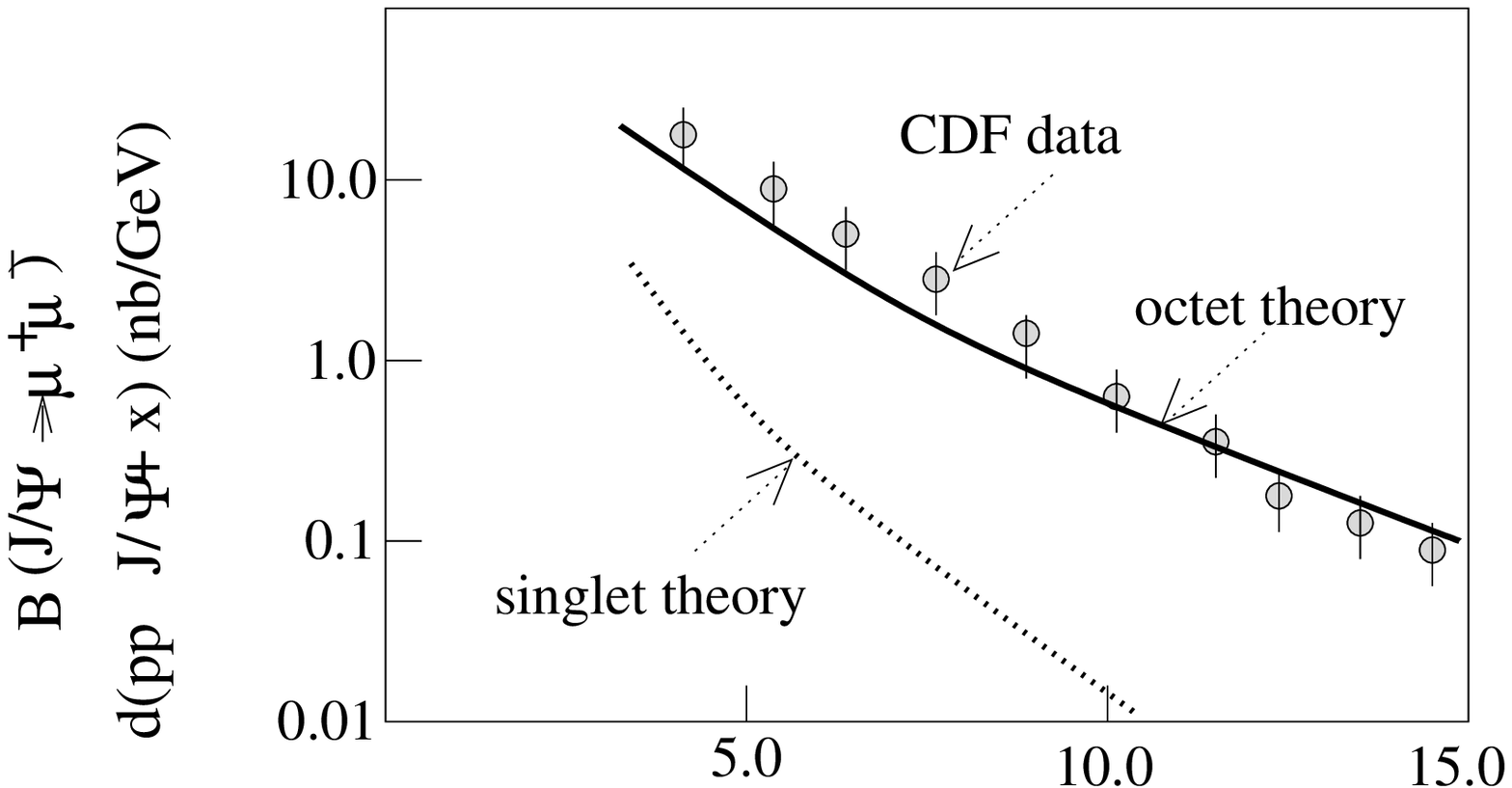,height=8cm,width=10cm}
\caption{d$\sigma$(pp$\rightarrow$J/$\Psi$)}
%\label{}
\end{center} 
\end{figure}

%\vspace{-2cm}
\subsection{ Proton-Proton Collisions and Production of $\Psi$ and
$\Upsilon$ States}

 In this subsection we review the publication of Ref\cite{klm11} on heavy 
quark state production in p-p collisions.
We only consider unpolarized p-p collisions. The production cross sections
are obtained from 
\beq
\label{ns1}
 \sigma_{pp\rightarrow \Phi(\lambda)} &=& \int_a^1  \frac{d x}{x}
f_q(x,2m)f_{\bar{q}}(a/x,2m) \sigma_{q\bar{q} \rightarrow \Phi(\lambda)} 
\nonumber \\
&& +f_g(x,2m)f_g(a/x,2m) \sigma_{g g \rightarrow \Phi(\lambda)} \; ,
\eeq
where $a= 4m^2/s$, with $m=1.5$  GeV for charmonium, and 5 GeV for 
bottomonium.
$f_g(x,2m)$, $f_q(x,2m)$ are the gluonic and quark distribution functions 
evaluated at $Q=2m$. 

For the quark and gluon cross sections, $\sigma_{q\bar{q} 
\rightarrow \Phi(\lambda)}$ and $\sigma_{g g \rightarrow \Phi(\lambda)}$ one needs
the octet matrix elements derived from the diagrams shown in Fig.10 by 
Braaten and Chen\cite{bc96}. The procedure of Nyyak and Smith\cite{ns06} was
followed in Ref\cite{klm11}.  The three octet matrix elements 
needed are $<O_8^\Phi(^1S_0)>$, $<O_8^\Phi(^3S_1)>$, and $<O_8^\Phi(^3P_0)>$,
with $\Phi$ either $J/\Psi$, $\Psi'(2S)$, or $\Upsilon(nS)$. Since these 
matrix elements are not well known, Nyyak and Smith\cite{ns06} use three 
scenerios:
\beq
\label{octetmatrixelements}
1)<O_8^\Phi(^1S_0)>&=& <O_8^\Phi(^3P_0)>/m^2=.0087, \nonumber \\
2)<O_8^\Phi(^1S_0)>&=&.039 {\rm \;and\;} <O_8^\Phi(^3P_0)> =0, \\
3)<O_8^\Phi(^1S_0)>&=&0 \; ,
\eeq
and  $<O_8^\Phi(^3P_0)>/m^2$=.01125, with $<O_8^\Phi(^3S_1)>$=.0112 in all
scenerios. All matrix elements have units GeV$^3$. Note that these matrix 
elements are not used to obtain the wave functions of the heavy quark meson 
states.

\newpage

  Using\cite{ns06} scenerio 2 the production cross sections\cite{bc96,ns06} 
for $\Phi$ for helicity $\lambda$ = 0 and 1 are

\beq
\label{spptoPhi}
  \sigma_{pp\rightarrow \Phi(\lambda=0)} &=& A_\Phi \int_a^1 \frac{d x}{x} 
f_g(x,2m)f_g(a/x,2m) \nonumber \\
\sigma_{pp\rightarrow \Phi(\lambda=1)} &=& A_\Phi \int_a^1 \frac{d x}{x}
[f_g(x,2m)f_g(a/x,2m)+0.613((f_d(x,2m)f_{\bar{d}}(a/x,2m) \nonumber \\
    &&+f_u(x,2m)f_{\bar{u}}(a/x,2m))]
\; ,
\eeq
with $A_\Phi=\frac{5 \pi^3 \alpha_s^2}{288 m^3 s}<O_8^\Phi(^1S_0)>$.

  The main purpose of this work was to explore the effects of matrix
elements for $\Psi'(2S)$ and $\Upsilon(3S)$, comparing results 
with the hybrid model to the standard model. In the standard model the states 
are (nS) quark anti-quark states, and the ratios of the matrix elements for
n greater than 1 is given by the squares of the wave functions. Note that
the basis for the octet model being used is the nonrelatavistic QCD 
model\cite{cl96,bc96,fl96}, with a model potential for the quark anti-quark
interaction giving bound states. A harmonic oscillator potential can
be used to approximately give the energies of the first few states, which is 
what is needed in the present work. For the octet matrix elements the results
of Refs.\cite{cl96,bc96,fl96,ns06} were used, as discussed above. 

To approximate the ratios of matrix elements in a 
nonrelativistic quark model for these heavy quark meson states harmonic 
oscillator wave functions were used\cite{merqm}, with $\Phi(1S)=
2 Exp[-r/a_o]/a_o^{3/2}$, $\Phi(2S)=\Phi(1S)(1-r/a_0)/2^{3/2}$, $\Phi(3S)=
\Phi(1S)(1-2r/3a_o +2r^2/27a_o^2)/3^{3/2}$. Defining N1= $\int |\Phi(2S)|^2$ 
divided by 
$\int |\Phi(1S)|^2$ for the 2S to 1S probability, and simillarly N2 for the 
3S to 1S probability, we find N1=0.039, N2=0.0064, N3=N2/N1=.16. This is a 
very rough estimate. Therefore, we use $A_{\Psi'(2S)}=0.039A_{J/\Psi (1S)}$,  
$A_{\Upsilon(2S)}=0.039 A_{\Upsilon (1S)}$, and $A_{\Upsilon(3S)}=0.0064 A_{\Upsilon (1S)}$ 
in the standard model.

On the other hand in the mixed hybrid study both $\Psi'(2S)$
and $\Upsilon(3S)$ were found to be approximately 50\% hybrids. In 
Ref\cite{lsk09} it was shown, using the external field method, that the octet 
to singlet matrix element was enhanced by a factor of $\pi^2$ compared 
to the standard model, as illustrated in Fig.14. For mixed hybrids 
an enhancement factor of 3.0 was used.
\vspace{5mm}
\begin{figure}[ht]
\begin{center}
\epsfig{file=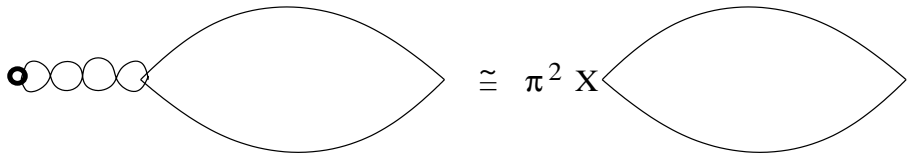,height=2.cm,width=12cm}
\caption{External field method for $\Psi'(2S)$ and $\Upsilon(3S)$ states}
\label{Figure 3}
\end{center}
\end{figure}
\newpage

  For differential cross sections the rapidity variable, $y$, is used,
\beq
\label{2}
      y(x) &=& \frac{1}{2} ln (\frac{E + p_z}{E-p_z}) {\rm ;\;with\;} 
E= \sqrt{M^2 + p_z^2}  \nonumber \\
             p_z &=& \frac{\sqrt{s}}{2} (x-\frac{a}{x})  \; ,
\eeq
or
\beq
\label{x(y)}
      x(y) &=& 0.5 \left[\frac{m}{s}(\exp{y}-\exp{(-y)})+\sqrt{(\frac{m}{s}
(\exp{y}-\exp{(-y)}))^2 +4a}\right]
\eeq
   For the unpolarized proton collisions we use a polynomial fit to the parton 
distributions of Ref.\cite{CTEQ6}. Because of the wide range of vaues, in 
order to obtain a good polynomial fit to the parton distributions we limit 
the range of rapidity to $-1. < y <1.$

   For Q=3 GeV, with m=Charmonium mass = 1.5 GeV, from Eq(\ref{x(y)}), x has a
range about 0.028 to 0.032, and a/x 0.008 to 0.015.  In Ref\cite{klm11} the 
following expressions were derived for the gluon (g), u and d quark, and 
anti-quark  distribution functions using QTEQ6 for Q=3 GeV, fitting the range 
x=0.008 to .004, which is needed for $\sqrt{s} \simeq$ 200 to 500 GeV
\beq
\label{3}
      f_g(x) & \simeq & 1334.21 - 67056.5 x + 887962.0 x^2  \nonumber \\
      f_d(x) & \simeq &72.956 - 3281.1 x + 42247.6 x^2 \nonumber \\
      f_u(x) & \simeq & 82.33 - 3582.36 x + 45867.3 x^2 \nonumber \\
  f_{\bar{u}}(x) & \simeq &55.98 - 2722.04 x + 35641.2 x^2 \\  
  f_{\bar{d}}(x) & \simeq &57.44 - 2757.05 x + 36030.5 x^2 \nonumber \; .
\eeq
 
 For Q=10 GeV, m=Bottomonium mass=5 GeV, from Eq(\ref{x(y)}), x has a
range about 0.05 to 0.08, and a/x 0.03 to 0.05. We have derived the 
following expressions for the gluon (g), u and d quark, and antiquark 
distribution functions using QTEQ6 for Q=10 GeV, fitting the range x=0.03 to 
.08, which is needed for $\sqrt{s}$=38.8 GeV and 2.76 TeV.
\beq
\label{4}
      f_g(x) & \simeq & 275.14 - 6167.6 x + 36871.3 x^2 \nonumber \\
      f_d(x) & \simeq & 26.96 - 527.14 x + 3119.13 x^2 \nonumber \\
      f_u(x) & \simeq & 32.92 - 604.38 x + 3530.1 x^2 \nonumber \\
  f_{\bar{u}}(x) & \simeq & 16.64 - 377.53 x + 2336.86 x^2 \\  
  f_{\bar{d}}(x) & \simeq & 17.81 - 390.64 x + 2392.46 x^2 \nonumber \; .
\eeq

The differential rapidity distribution for $\lambda=0$ is given by
\beq
\label{5}
      \frac{d \sigma_{pp\rightarrow \Phi(\lambda=0)}}{dy} &=& 
     A_\Phi \frac{1}{x(y)} f_g(x(y),2m)f_g(a/x(y),2m) \frac{dx}{dy} \; ,
\eeq

while for $\lambda$=1
\beq
\label{6}
\frac{d \sigma_{pp\rightarrow \Phi(\lambda=1)}}{dy} &=& A_\Phi \frac{1}{x(y)}
[f_g(x(y),2m)f_g(a/x(y),2m)+0.613(f_d(x(y),2m)f_{\bar{d}}(a/x(y),2m) 
\nonumber \\
    &&+f_u(x(y),2m)f_{\bar{u}}(a/x(y),2m)]\frac{dx}{dy} \; .
\eeq

\clearpage
 \subsubsection{Charmonium  Production Via Unpolarized p-p
Collisions at E=$\sqrt{s}$= 200 GeV at BNL-RHIC}

 Unpolarized p-p collisions for $\sqrt{s}=200 GeV$  
corresponding to BNL energy, using scenerio, with the 
nonperturbative matrix elements given above, $A_\Phi=
\frac{5 \pi^3 \alpha_s^2}{288 m^3 s} <O_8^\Phi(^1S_0)>$ =$7.9 \times 10^{-4}$nb 
for $\Phi$=$J/\Psi$ and $2.13 \times  10^{-5}$nb for $\Upsilon(1S)$ heavy 
quark states. 

For  $\sqrt{s}=200 GeV$ 
\beq
\label{7}
   x(y) &=& 0.5 \left[\frac{m}{200}(\exp{y}-\exp{(-y)})+\sqrt{(\frac{m}{200}
(\exp{y}-\exp{(-y)}))^2 +4a}\right] \nonumber \\
  \frac{d x(y)}{d y} &=&\frac{M}{400}(\exp{y}+\exp{(-y)})\left[1. + 
\frac{\frac{M}{200}(\exp{y}-\exp{(-y)})}{\sqrt{(\frac{M}{200} 
(\exp{y}-\exp{(-y)}))^2 +4a}}\right] \; .
\eeq
Note that there was a typo error in Ref\cite{klm11}, with $\frac{M}{200}
(\exp{y}+\exp{(-y)})$ instead of $\frac{M}{200}(\exp{y}-\exp{(-y)})$ in
the numerator of Eq(\ref{7}). 
  Using Eqs(\ref{5},\ref{6},\ref{7}), with the parton distribution functions 
given in Eq(\ref{3}), we find d$\sigma$/dy for Q=3 GeV, $\lambda=0$ and 
$\lambda=1$ the results for $J/\Psi$ shown in  Figure  15.

\begin{figure}[ht]
\begin{center}
\epsfig{file=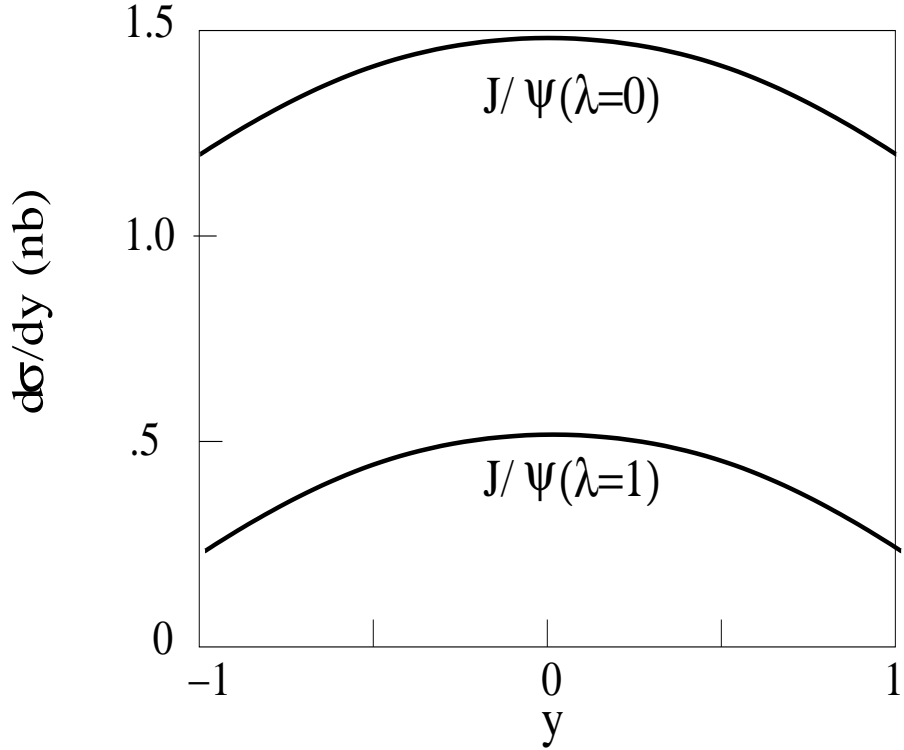,height=10 cm,width=12cm}
\caption{d$\sigma$/dy for Q=3 GeV, E=200 GeV unpolarized p-p collisions 
producing $J/\Psi$
with $\lambda=0$, $\lambda=1$}
%\label{}
\end{center}
\end{figure}

Note that the shape of d$\sigma$/dy is consistent with the BNL-RHIC-PHENIX
detector rapidity distribution\cite{cln04}.
\clearpage

For $\Psi'(2S)$ the results are shown in Figure 16 for both the standard
model and the mixed hybrid theory.
\vspace{6cm}

\begin{figure}[ht]
\begin{center}
\epsfig{file=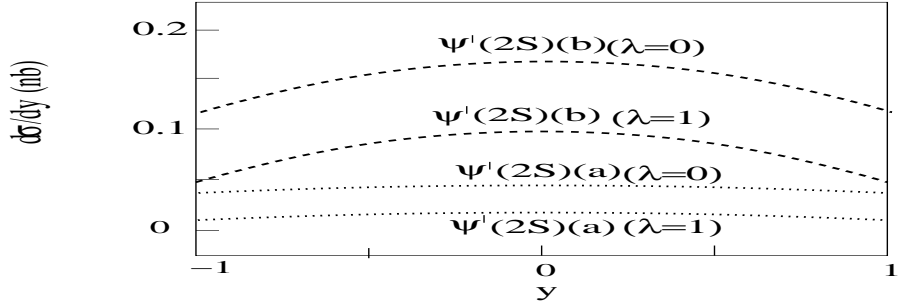,height=4cm,width=12cm}
\caption{d$\sigma$/dy for Q= 3 GeV, E=200 GeV unpolarized p-p  collisions 
producing$\Psi'(2S)$
 with $\lambda=1$,$\lambda=0$}
%\label{}
\end{center}
\end{figure}

The results for d$\sigma$/dy shown in Figure 16 labeled $\Psi'(2S)(a)$ are
 obtained by using for the standard nonperturbative matrix element=0.039 
times the matrix elements for $J/\Psi$ production; while the results labeled 
$\Psi'(2S)(b)$ are obtained by using
the matrix element derived using the result that the $\Psi'(2S)$ is 
approximately 50\% a hybrid with the enhancement is at least a factor of $\pi$,
as discussed above.

\subsubsection{Upsilon Production Via Unpolarized p-p Collisions at 
E=$\sqrt{s}$= 38.8 GeV at Fermilab}

In this subsection the cross sections calculated for
for $\Upsilon(nS)$ production, with n= 1, 2, 3  at 38.8, which has
been measured at Fermilab\cite{fermi91,fermi94}, are reviewed. 

 For Q=10 GeV, using the parton distributions given in Eq(\ref{4}) and 
Eqs(\ref{5},\ref{6}) for helicity $\lambda=0$, $\lambda=1$, with
$A_{\Upsilon}$ =$5.66 \times 10^{-4}$nb and $a=6.64 \times 10^{-2}$, one obtains
$d\sigma/dy$ for $\Upsilon(nS)$ production.

\clearpage

 The results for $\Upsilon(1S)$, $\Upsilon(2S)$ are shown in Figure 17, 
and for $\Upsilon(3S)$ in Figure 18.
\vspace{7.5cm}

\begin{figure}[ht]
\begin{center}
\epsfig{file=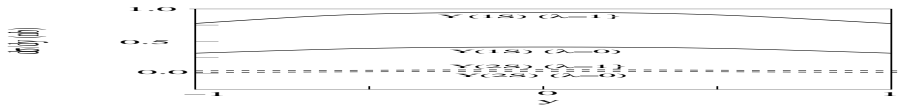,height=1.3cm,width=12cm}
\caption{d$\sigma$/dy for Q= 10 GeV, E=38.8 GeV unpolarized p-p collisions 
producing $\Upsilon(1S)$, $\Upsilon(2S)$ with $\lambda=0$, $\lambda=1$}
\label{}
\end{center}
\end{figure}

\vspace{5.5cm}

\begin{figure}[ht]
\begin{center}
\epsfig{file=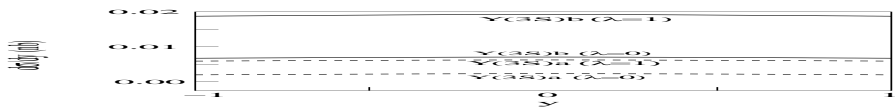,height=1.3cm,width=12cm}
\caption{d$\sigma$/dy for Q= 10 GeV, E=38.8 GeV unpolarized p-p collisions 
producing $\Upsilon(3S)$ with $\lambda=0$, $\lambda=1$. Results labeled a,b
are for the standard, mixed hybrid theories.}
\label{}
\end{center}
\end{figure}

\newpage

 It should be noted that the ratios of $d\sigma/dy$ for $\Psi'(2S)$/$J/\Psi$,
and $\Upsilon(3S)$/($\Upsilon(1S)$+$\Upsilon(2S)$) for the hybrid theory vs. 
the standard are our most significant results, as there are uncertainties in 
the absolute magnitudes and shapes of $d\sigma/dy$ on the scenerios, as
well as the magnitudes of the matrix elements.

\subsubsection{Polarized p-p collisions for E=200 GeV at BNL-RHIC}

For polarized p-p collisions the equations for  
$\frac{d \sigma_{pp\rightarrow \Phi(\lambda=0)}}{dy}$ and
$ \frac{d \sigma_{pp\rightarrow \Phi(\lambda=1)}}{dy}$ are the same as 
Eqs(\ref{5},\ref{6}) with the parton distribution functions $fg$ and $fq$
given in Eqs(\ref{3},\ref{4}) replaced by $\Delta fg$ and $\Delta fq$,
the parton distribution functions for longitudinally polarized p-p collisions.
A fit to the parton distribution functions for polarized p-p collisions
for Q=3 GeV obtained from CTEQ6\cite{CTEQ6} in the x range needed for
$\sqrt{s}$=200 GeV is
\beq
\label{8}
\Delta f_g(x) &\simeq & 15.99-700.34 x+13885.4 x^2-97888. x^3 \nonumber \\
\Delta f_d(x) & \simeq & -5.378.+205.60 x-4032.77 x^2+28371. x^3 \nonumber \\
\Delta f_u(x) & \simeq & 8.44-292.19 x+5675.16 x^2-39722. x^3 \nonumber \\
\Delta f_{\bar{u}}(x) & \simeq &-1.447 +64.67 x-1268.24 x^2+8878.32 x^3  \\  
  \Delta f_{\bar{d}}(x) &=&\Delta f_{\bar{u}}(x) \nonumber \; ,
\eeq
and for Q=10 GeV, which we do not use in the present work, as the 
$\Upsilon(nS)$ are not resolved at BNL-RHIC,

\beq
\label{9}
\Delta f_g10(x) & \simeq & 28.98-1435.47 x+29533.5 x^2-211440. x^3 \nonumber \\
\Delta f_d10(x) & \simeq &-6.074+241.57 x-4762.04 x^2+33604.4 x^3 \nonumber \\
\Delta f_u10(x) & \simeq & 9.88-348.632 x+6729.49 x^2-47058. x^3  \nonumber \\
\Delta f_{\bar{u}}10(x) & \simeq & -1.552+75.731 x-1531.97 x^2+10896.6 x^3  \\  
\Delta f_{\bar{d}}10(x) &=& \Delta f_{\bar{u}}10(x) \nonumber \; .
\eeq

The differential rapidity distribution for polarized p-p collisions are
\beq
\label{Deltasig0}
      \frac{d \Delta \sigma_{pp\rightarrow \Phi(\lambda=0)}}{dy} &=& 
  - A_\Phi \frac{1}{x(y)}\Delta f_g(x(y),2m)\Delta f_g(a/x(y),2m) \frac{dx}{dy} 
\; ,
\eeq

\beq
\label{Deltasig1}
\frac{d \Delta \sigma_{pp\rightarrow \Phi(\lambda=1)}}{dy} &=& -A_\Phi 
\frac{1}{x}[\Delta f_g(x(y),2m) \Delta f_g(a/x(y),2m)-0.613(\Delta f_d(x(y),2m) 
\nonumber \\
 &&\Delta f_{\bar{d}}(a/x(y),2m) +\Delta f_u(x(y),2m)
\Delta f_{\bar{u}}(a/x(y),2m)]\frac{dx}{dy} \; .
\eeq

For polarized p-p collisions, Q=3 GeV,  the results for d$ \Delta \sigma$/dy for
$J/\Psi$ production using the standard model are shown in Figure 19 
while for $\Psi'(2S)$ the results are shown in Figure 20.  As above, the 
curves labelled $\Psi'(2S)a$ and are the standard model results, while that 
labelled $\Psi'(2S)b$ are the results for a mixed hybrid.
The enhancement from active glue is once more quite evident.
Since $\Upsilon(nS)$ states have not been resolved at BNL-RHIC, where polarized
p-p collisions were measured, we do not calculate d$ \Delta \sigma$/dy for 
$\Upsilon(nS)$ states.

\clearpage

\begin{figure}[ht]
\begin{center}
\epsfig{file=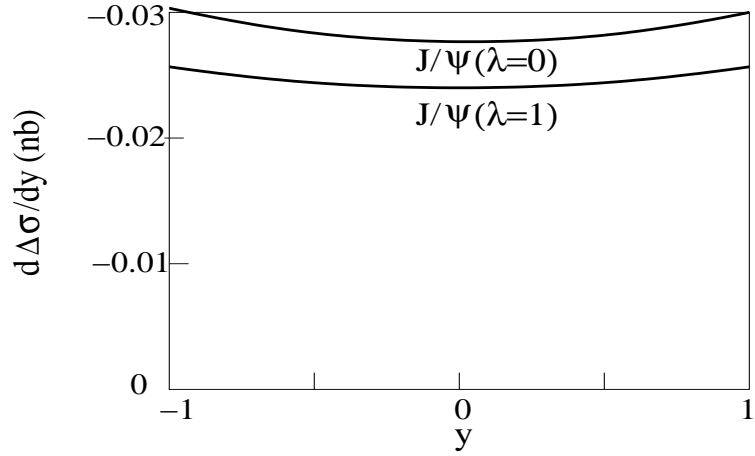,height=6cm,width=10cm}
\caption{d$ \Delta \sigma$/dy for Q=3 GeV, E=200 GeV polarized p-p collisions 
producing $J/\Psi$, with $\lambda = 0$, $\lambda = 1$}
%\label{}
\end{center}
\end{figure}

\begin{figure}[ht]
\begin{center}
\epsfig{file=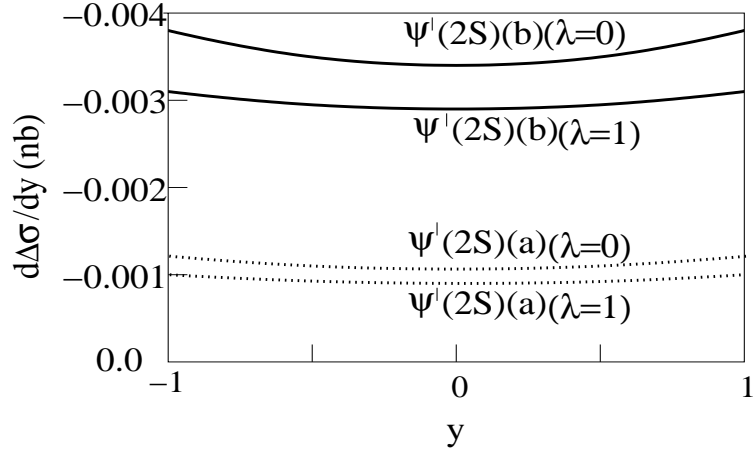,height=6cm,width=10cm}
\caption{d$ \Delta \sigma$/dy for Q= 3 GeV, E=200 GeV polarized p-p  collisions 
producing $\Psi'(2S)$
with $\lambda=0$, $\lambda=1$}
%\label{}
\end{center}
\end{figure}

Once again, it is the ratios of $d \Delta \sigma/dy$ that are 
most significant, as there is uncertainty both the absolute magnitudes and
shapes.
\clearpage

 \subsubsection{Ratios of Cross Sections for $\Psi$,  $\Upsilon$  
Production Via p-p Collisions }

Because of problems with normalization we cannot compare our cross sections
directly with experiment, but a comparison of ratios of cross sections with 
experiment is an excellent test of the theory
used to estimate $\Psi'(2S), J/\Psi(1)$ $\Psi$ and $\Upsilon$  production.

In this subsection the cross sections for $\Psi'(2S), J/\Psi(1)$, 
$\Upsilon(nS)$,  production, with 
n= 1, 2, 3 are calculated, and then the theory that $\Psi'(2S), \Upsilon(3S)$
are hybrids 
is used to estimate the ratios of cross section. Since
with scenerio 2 with $<O_8^\Phi(^3P_0)>$=0, the $\lambda=0$ helicity
dominates the cross section\cite{ns06},  the $\lambda=1$ terms were dropped.  
From Eq(\ref{spptoPhi}), for $\lambda=0$, the cross section is determined from
\beq
\label{sigmapp}
  \sigma_{pp\rightarrow \Phi(\lambda=0)} &=& A_\Phi \int_a^1 \frac{d x}{x} 
f_g(x,2m)f_g(a/x,2m) \; ,
\eeq
where 
\beq
\label{APhi}
  A_\Phi & \propto& \frac{1}{s} \; ,
\eeq
with $s=E^2$, as discussed above. The energy dependence of 
$\sigma_{pp\rightarrow \Phi(\lambda=0)}$ of Eq(\ref{sigmapp}), given by Eq(\ref{APhi})
will be compared to experiment in the next subsection. 

As discussed ref\cite{klm11} the estimated
ratios for p-p production of $\Psi(2S)$ and $J/\Psi(1S)$ using the
harmonic-oscillator wave functions for the standard model and a
factor $\simeq \pi$ for the mixed hybrid theory are
\beq
\label{ppPsiratio}
    \sigma(\Psi(2S))/\sigma(J/\Psi(1S))|_{standard} &\simeq& 0.039 \nonumber \\
    \sigma(\Psi(2S))/\sigma(J/\Psi(1S))|_{hybrid} &\simeq& 0.122 \; ,
\eeq
while the estimated
$\Upsilon(2S),\Upsilon(3S)$ to $\Upsilon(1S)$ ratios are
\beq
\label{ppUpratio}
    \sigma(\Upsilon(2S))/\sigma(\Upsilon(1S))|_{standard} &\simeq& 
\sigma(\Upsilon(2S))/\sigma(\Upsilon(1S))|_{hybrid} \simeq 0.039 \nonumber \\
 \sigma(\Upsilon(3S))/\sigma(\Upsilon(1S))|_{standard} &\simeq& .0064 
\nonumber \\
 \sigma(\Upsilon(3S))/\sigma(\Upsilon(1S))|_{hybrid} &\simeq& 0.0201 \; .
\eeq

  From the recent measurements by the ALICE Collaboration\cite{alice14}
 the $\Psi(2S)$ to $J/\Psi(1S)$ ratio is
\beq
\label{alicepsi}
   \frac{\sigma_{\Psi(2S)}}{\sigma_{J/\Psi}} \simeq 0.170 \pm 0.011(stat)
\pm 0.013(syst) \; ,
\eeq
one can see from Eq(\ref{alicepsi}) that the $ \sigma_{\Psi(2S)}/\sigma_{J/\Psi}$
ratio is much larger than the standard model and is consistent with the 
mixed hybrid theory\cite{lsk09} within theoretical and expermental errors. 

  The $\Upsilon(2S),\Upsilon(3S)$ to $\Upsilon(1S)$ ratios are difficult for
experiments to measure. These ratios at E= 7 TeV
were recently measured by the ATLAS Collaboration\cite{atlas13}. The
ratios of cross sections include the branching fractions, $BR(\Upsilon 
\rightarrow \mu^+ \mu^-)$, with the experimental results
Because of the branching fractions, it is difficult to  compare the
ATLAS results to the theoretical cross section ratios  given
in Eq(\ref{ppUpratio}). However, the energy dependence of $\Upsilon$ cross
section can be measured, as discussed in the next subsection.
\newpage

\subsubsection{Theoretical vs Experimental Energy Dependence of $\Upsilon$
Cross Sections}

    Note that $A_\Phi$ from Eq(\ref{APhi}) has the property 
$A_\Phi(s) \propto 1/s$, so cross sections should also be $\propto 1/s$.
Recently, LHCb measured experimental ratios at 7 and 8 TeV at forward 
rapidity for $\Upsilon(1S),\Upsilon(2S),\Upsilon(3S)$ production\cite{LHCb15}.
The theoretical and the experiment ratios are
\beq
\label{Eratio}
   (\sigma_{\Upsilon}(8 TeV)/\sigma_{\Upsilon}(7 TeV))_{theory} & \simeq & 
1.306 \nonumber \\
    (\sigma_{\Upsilon}(8 TeV)/\sigma_{\Upsilon}(7 TeV))_{experiment} & \simeq &
 1.291 \pm 0.005
\; ,
\eeq
so the theoretical ratio for different energies is consistent with experiment
within errors.

\subsubsection{Conclusions of Ref\cite{klm11}}

   The mixed hybrid theory for heavy quark states was used to 
predict that the cross sections for production of the charmonium $\Psi'(2S)$
state in 200 GeV p-p collisions and bottomonium $\Upsilon(3S)$ states in 38.8
 GeV p-p collisions are much larger than the standard model. Also the 
estimated ratio of cross sections for 2.76 TeV and 38.8 GeV experiments, and 
the prediction for the $\Upsilon(3S)$ production cross section is larger 
than the standard model, and closer to the experimental values. 

  Because of the importance of gluonic production in processes in 
a Quark Gluon Plasma, this could lead to a test of the creation of 
QGP in RHIC. 

\subsubsection{Upsilon Production In p-p Collisions For Forward Rapidities At 
LHC}

  In the work on p-p collisions producing heavy quark states reviewed above
the rapidity was y=-1 to 1,  while the present study is for y=2.5 to 4.0 at 
the LHC\cite{kd13}.  The differential rapidity distribution for Upsilon 
production with  $\lambda=0$ (dominant for $\Upsilon(nS)$ production), as 
is given by
\beq
\label{ppUpsilon}
      \frac{d \sigma_{pp\rightarrow \Phi(\lambda=0)}}{dy} &=& 
     A_\Upsilon \frac{1}{x(y)} f_g(x(y),2m)f_g(a/x(y),2m) \frac{dx}{dy} \; ,
\eeq 
with $x(y),\frac{dx(y)}{dy}$ defined in Eq(\ref{7}) and $A_\Upsilon=1.12\times 
10^{-7},1.73\times 10^{-8}$ nb, for $\sqrt{s}$ = 2.76, 7.0 TeV. $f_g$  is the 
gluonic distribution function given in Eq(\ref{4}) for the energies at
the LHC.

    Using Eqs(\ref{ppUpsilon},\ref{4}) and parameters given in 
Ref\cite{klm11} we obtain
the results for $\Upsilon(1S)$ and $\Upsilon(3S)$ production shown in Fig. 21
and Fig. 22 at 2.76 TeV and 7.0 TeV\cite{k12} in p-p collisions for $2.5 \leq y 
\leq 4.0$.  Although the units in Figs. 21, 22 are in pb, the actual 
magnitude is uncertain due to the normalization of the state. The overall 
magnitude and rapidity dependence of the differential rapidity distribution,
 however, provides satisfactory estimates at forward rapidities for LHC 
experiments.

   Also, it is the ratios of cross sections, 
$\sigma(\Upsilon(2S))/\sigma(\Upsilon(1S))$ and $\sigma(\Upsilon(3S))/
\sigma(\Upsilon(1S))$ which are most accurate, and are used to prove that
the mixed hybrid theory for $\Upsilon(3S))$ is much better than the standard
$b\bar{b}$ model. This is discussed in detail in the next section.

\clearpage

\begin{figure}[ht]
\begin{center}
\epsfig{file=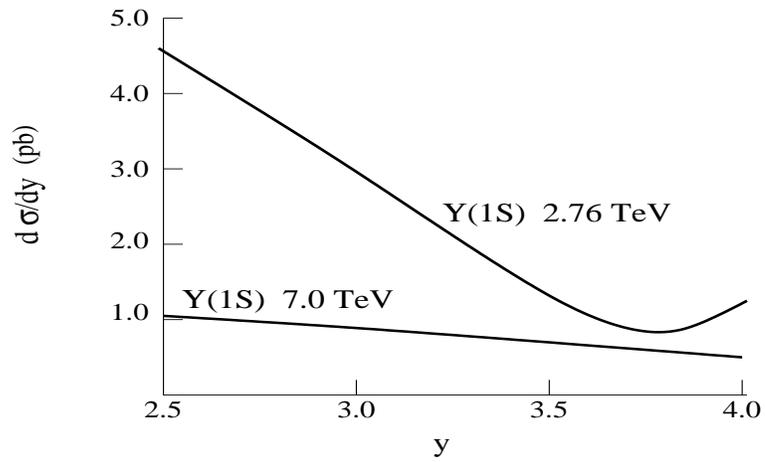,height=6cm,width=10cm}
\caption{d$\sigma$/dy for pp collisions at $\sqrt{s}$ = 2.76 and 7.0 
producing $\Upsilon(1S)$.}
\label{Figure 18}
\end{center}
\end{figure}
\vspace{2cm} 

\begin{figure}[ht]
\begin{center}
\epsfig{file=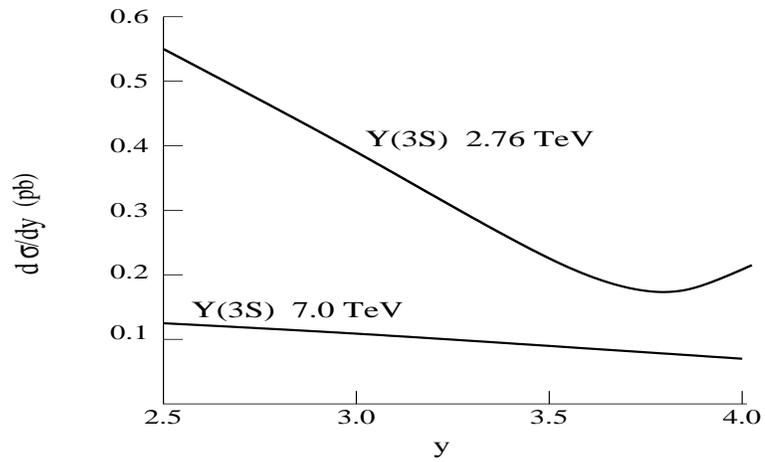,height=6cm,width=10cm}
\caption{d$\sigma$/dy for pp collisions at $\sqrt{s}$ = 2.76 and 7.0 TeV 
producing $\Upsilon(3S)$.}
\label{Figure 19}
\end{center}
\end{figure} 
\newpage

\subsubsection{$\Psi$ and $\Upsilon$ Production In pp Collisions at E=7.0 TeV}

This is an extension of recent studies for $\Upsilon(nS)$ and $\Psi(1S,2S)$ 
production at the LHC in p-p collisions with E=7.0 GeV and the ALICE 
detector\cite{kd213}. The differential rapidity cross section is the same as 
Eq(\ref{ppUpsilon}) with  $A_\Upsilon =1.73\times 10^{-8}$ nb for E= 7.0 TeV, and
$A_\Upsilon \rightarrow A_\Psi=6.46 \times 10^{-7}$. The gluonic distribution 
$f_g$ is the same as in Eq(\ref{4}). The calculation of the production of $\Upsilon(3S)$ and $\Psi(2S)$ states is done with the mixed heavy hybrid 
theory\cite{lsk09}.
\vspace{5cm}

\begin{figure}[ht]
\begin{center}
\epsfig{file=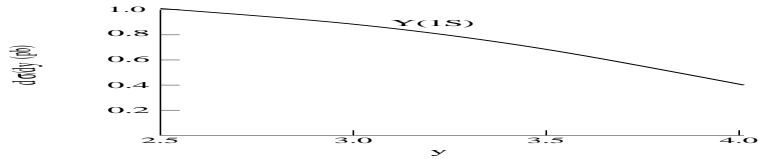,height=2cm,width=10cm}
\caption{d$\sigma$/dy for pp collisions at $\sqrt{s}$ = 7.0 TeV 
producing $\Upsilon(1S)$.}
\label{}
\end{center}
\end{figure} 

\clearpage

The differential rapidity cross sections for $\Upsilon(1S)$, $\Upsilon(2S)$ 
with the standard model are shown in Figs. 23, 24; and for $\Upsilon(3S)$ 
with the standard model and mixed hybrid theory 
are shown in Fig. 25.
\vspace{3cm} 

\begin{figure}[ht]
\begin{center}
\epsfig{file=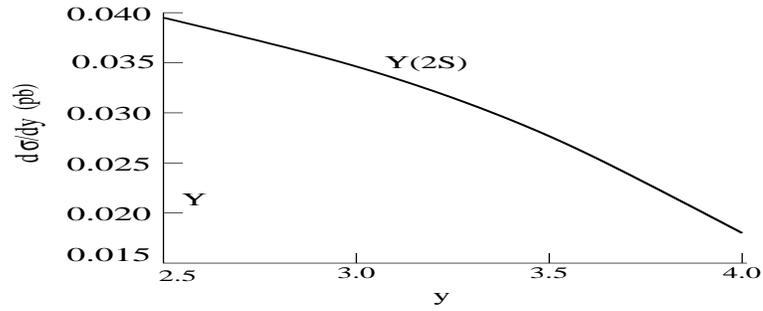,height=4cm,width=10cm}
\caption{d$\sigma$/dy for pp collisions at $\sqrt{s}$ = 7.0 TeV 
producing $\Upsilon(2S)$.}
\label{}
\end{center}
\end{figure} 
\vspace{5mm}
 
\vspace{2cm} 

\begin{figure}[ht]
\begin{center}
\epsfig{file=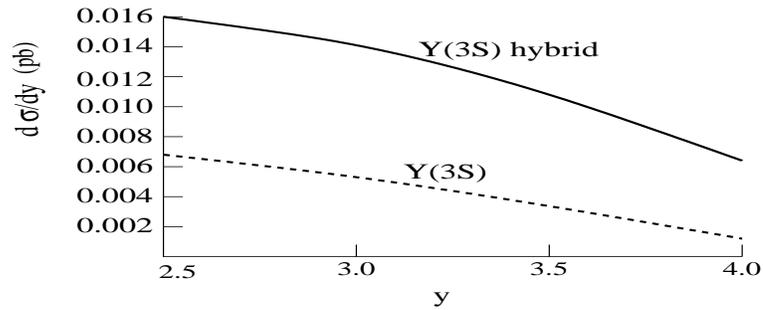,height=4cm,width=10cm}
\caption{d$\sigma$/dy for pp collisions at $\sqrt{s}$ = 7.0 TeV 
producing $\Upsilon(3S)$ for usual and hybrid theories.}
\label{Figure 2}
\end{center}
\end{figure} 
\clearpage
 The differential rapidity cross sections for $J/\Psi(1S)$ and $\Psi(2S)$
are shown in Figures 26 and 27.
\vspace{3cm}
\begin{figure}[ht]
\begin{center}
\epsfig{file=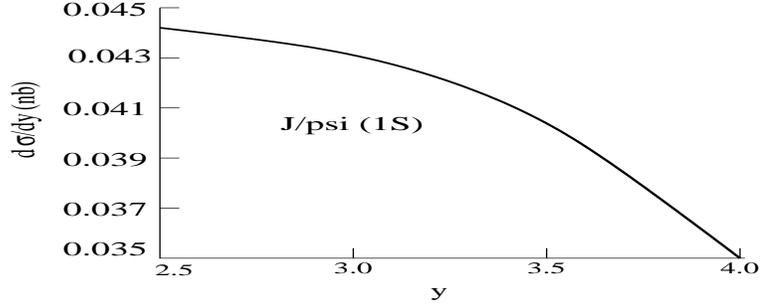,height=4cm,width=10cm}
\caption{d$\sigma$/dy for pp collisions at $\sqrt{s}$ = 7.0 TeV 
producing $J/\Psi (1S)$.}
\label{}
\end{center}
\end{figure}
\vspace{4cm}

\begin{figure}[ht]
\begin{center}
\epsfig{file=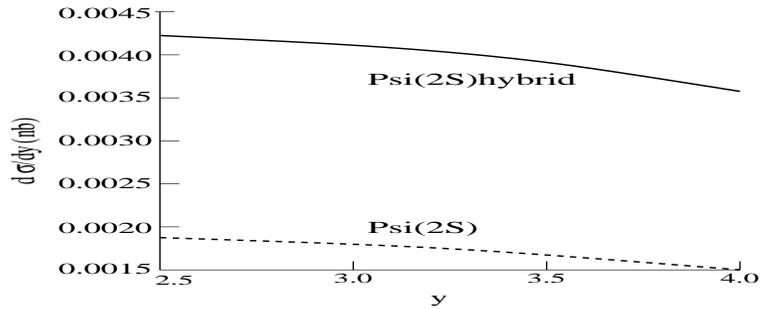,height=4cm,width=10cm}
\caption{d$\sigma$/dy for pp collisions at $\sqrt{s}$ = 7.0 TeV 
producing $\Psi (2S)$ for usual and hybrid theories.}
\label{}
\end{center}
\end{figure} 
\vspace{-5mm}

For $\Upsilon(3S)$ and $\Psi(2S)$ the standard $q\bar{q}$ prediction is
shown by dashed curves, while the prediction using the mixed hybrid 
theory\cite{lsk09} is shown with solid curves, with the difference
explained in Ref\cite{klm11}.
\clearpage

\subsubsection{$\Psi$ and $\Upsilon$ Production In p-p Collisions at E=8.0 TeV}

  This is an extension of the preceeding subsubsection for $\Upsilon(nS), 
n=1,2,3,$ and $J/\Psi(1S),\Psi(2S)$ production in p-p collisions with
the ALICE detector at 7.0 TeV, with new predictions for p-p collisions at
the LHC-ALICE with E=8.0 TeV\cite{kd14}. The differential rapidity cross 
section is the same as Eq(\ref{ppUpsilon}) with $A_\Upsilon =1.33 \times 10^{-8}$ 
and $A_\Upsilon\rightarrow A_\Psi =4.95 \times 10^{-7}$ for E= 8.0 TeV. The gluonic 
distribution $f_g(x(y),2m)$ for the range of $x$ needed for $E=8.0$ TeV is the
same as Eq(\ref{4}).

The calculation of the production of $\Upsilon(3S)$ and $\Psi(2S)$ states
is done with the usual quark-antiquark model and the mixed heavy quark
hybrid theory, as in the previous subsections. 

The differential rapidity cross sections for $J/\Psi(1S)$ and $\Psi(2S)$  
production for the standard model and the mixed hybrid theory are shown in 
Figure 28.
\vspace{8cm} 

\begin{figure}[ht]
\begin{center}
\epsfig{file=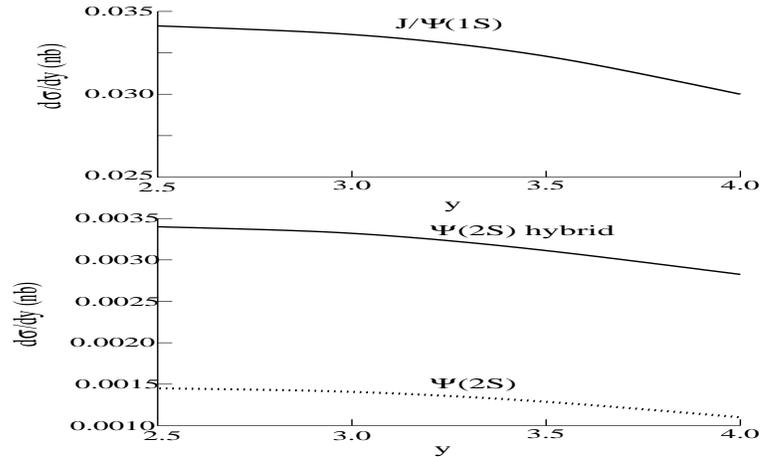,height=6cm,width=10cm}
\end{center}
\caption{d$\sigma$/dy for p-p collisions at $\sqrt{s}$ = 8.0 TeV 
producing $J/\Psi(1S)$; and $\Psi(2S)$ for the standard model (dashed curve) 
and the mixed hybrid theory.}
\label{Figure 20}
\end{figure} 
\clearpage

The differential rapidity cross sections for $\Upsilon(1S)$, $\Upsilon(2S)$, 
and $\Upsilon(3S)$ are shown in Figure 29.
\vspace{9cm}

\begin{figure}[ht]
\begin{center}
\epsfig{file=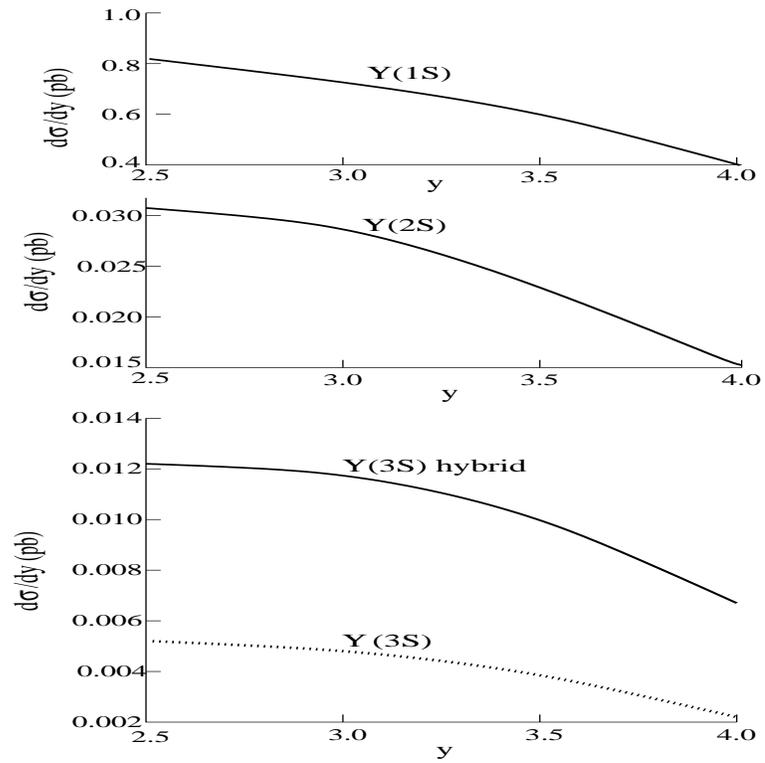,height=10cm,width=10cm}
\end{center}
\caption{d$\sigma$/dy for p-p collisions at $\sqrt{s}$ = 8.0 TeV for
producing $\Upsilon(1S)$, $\Upsilon(2S)$, $\Upsilon(3S)$ (dashed curve)
using the standard model; and $\Upsilon(3S)$ with the mixed hybrid theory.}
\label{Figure 21}
\end{figure} 
\clearpage

\subsubsection{$\Psi$ and $\Upsilon$ Production In p-p Collisions at E=13 TeV}

Motivated by the LHC modification in 2015, this subsubsection is an extension 
of the preceeding subsubsections for p-p collisions at E=7.0, 8.0 TeV with 
predictions of $\Upsilon(nS), n=1,2,3,$ and $J/\Psi(1S),\Psi(2S)$ production 
in p-p collisions at 13 TeV\cite{kd15}

The differential rapidity cross sections for $J/\Psi(1S)$ and $\Psi(2S)$ 
production for the standard model and the mixed hybrid theory for 
p-p collisions at E=13 TeV are shown in Figure 30.
\vspace{11cm} 

\begin{figure}[ht]
\begin{center}
\epsfig{file=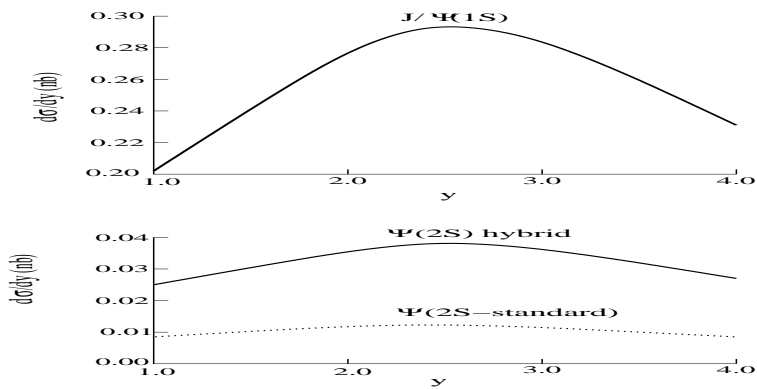,height=5cm,width=10cm}
\end{center}
\caption{d$\sigma$/dy for p-p collisions at $\sqrt{s}$ = 13.0 TeV 
producing $J/\Psi(1S)$; and $\Psi(2S)$ for the standard model (dashed curve) 
and the mixed hybrid theory.}
\label{}
\end{figure} 
\clearpage
Differential rapidity cross sections $\Upsilon$ production for p-p
collisions at 13 TeV are shown in Figure 31.
\vspace{16cm}

\begin{figure}[ht]
\begin{center}
\epsfig{file=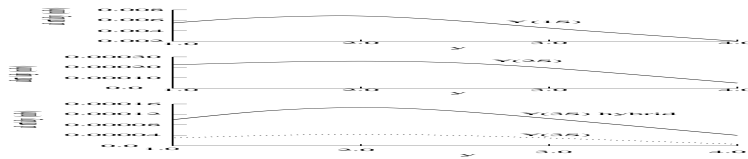,height=2cm,width=10cm}
\end{center}
\caption{d$\sigma$/dy for p-p collisions at $\sqrt{s}$ = 13.0 TeV 
producing $\Upsilon(1S)$, $\Upsilon(2S)$, and $\Upsilon(3S)$ for the standard 
model (dashed curve) and the mixed hybrid theory.}
\label{}
\end{figure}   
\clearpage

\subsubsection{$\Psi$ and $\Upsilon$ Production In p-p Collisions at E=14 TeV}

This subsubsection is an extension of the preceeding subsubsection for p-p 
collisions at E=13 TeV with predictions of  $J/\Psi(1S), \Psi(2S), \Upsilon(1S),
 \Upsilon(2S), \Upsilon(3S)$ production via p-p collisions at 14 TeV, based on
recent research \cite{kd215}. Although the rapidity 
dependence of d$\sigma$/dy, shown in the figures for p-p collisions at 14 TeV, 
are similar to those at 13TeV, with the LHC energy will be  increased to 14 TeV
during the LHC's second run period starting in 2015.
This should be useful for comparison with experiments. The differential 
rapidity cross sections for $J/\Psi(1S)$ and $\Psi(2S)$ production for the 
standard model and the mixed hybrid theory are shown in Figure 32.
\vspace{10cm} 

\begin{figure}[ht]
\begin{center}
\epsfig{file=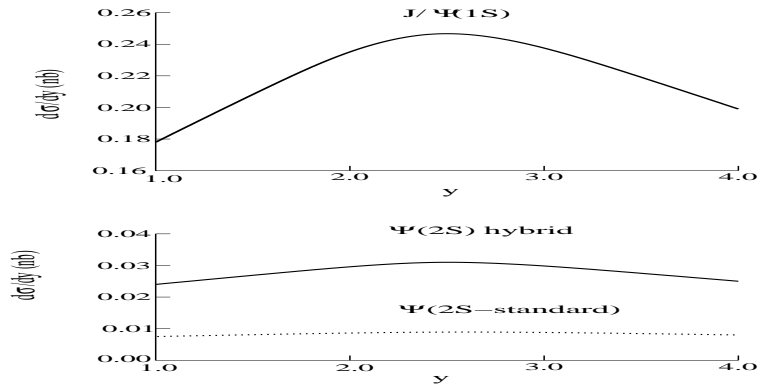,height=5cm,width=10cm}
\end{center}
\caption{d$\sigma$/dy for p-p collisions at $\sqrt{s}$ = 14.0 TeV 
producing $J/\Psi(1S)$; and $\Psi(2S)$ for the standard model (dashed curve) 
and the mixed hybrid theory.}
\label{Figure }
\end{figure}   

\clearpage

  The differential rapidity cross sections for $\Upsilon(1S),\Upsilon(2S),
 \Upsilon(3)$  production for the standard model and the mixed hybrid theory 
are shown in Figure 33.

\vspace{16cm}

\begin{figure}[ht]
\begin{center}
\epsfig{file=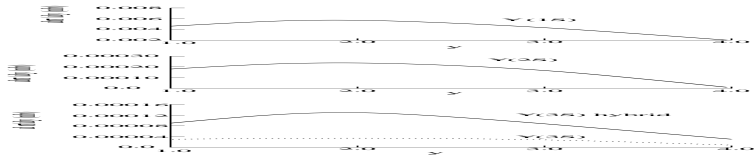,height=2cm,width=10cm}
\end{center}
\caption{d$\sigma$/dy for p-p collisions at $\sqrt{s}$ = 14.0 TeV 
producing $\Upsilon(1S)$, $\Upsilon(2S)$, and $\Upsilon(3S)$ for the standard 
model (dashed curve) and the mixed hybrid theory.}
\label{Figure }
\end{figure}   
  
\newpage

\section{Heavy-quark state production in A-A collisions
 at $\sqrt{s_{pp}}$=200 GeV }

 This section is a review of Ref\cite{klm14}. The differential rapidity cross 
section for the production of a heavy quark state $\Phi$ with helicirkty 
$\lambda=0$ in the color octet model via A-A collisions is given by

\beq
\label{dsAAlambda0}
   \frac{d \sigma_{AA\rightarrow \Phi(\lambda=0)}}{dy} &=& 
   R_{AA} N^{AA}_{bin}< \frac{d \sigma_{pp\rightarrow \Phi(\lambda=0)}}{dy}>
\; ,
\eeq
where $R_{AA}$ is the nuclear modification factor, defined in Ref\cite{phenix02},
which includes the dissociation factor after the state $\Phi$ is 
formed\cite{star02}. See Refs.\cite{vogt10},\cite{sv13} for a discussion of
 ``cold nuclear matter effects'' and references to earlier experimental and 
theoretical publications.  $N^{AA}_{bin}$ is the number of binary collisions in 
the A-A collision, and $< \frac{d \sigma_{pp\rightarrow \Phi(\lambda=0)}}{dy}>$ is the
differential rapidity cross section for $\Phi$ production via nucleon-nucleon
collisions in the nuclear medium. Note that $R^E_{AA}$, which
we take as a constant, can be functions of rapidity. See 
Refs\cite{vogt08,sv13} for a review and references to many publications.

Experimental studies show that for $\sqrt{s_{pp}}$ = 200 GeV
$R_{AA}\simeq 0.5$ both for Cu-Cu\cite{star09,phenix08} and Au-Au\cite{phenix07,
star07,kks06}. The number of binary collisions are  $N^{AA}_{bin}$=51.5 for 
Cu-Cu\cite{sbstar07} and 258 for Au-Au. The differential rapidity cross 
section for p-p collisions 
in terms of $f_g$\cite{CTEQ6,klm11}, the gluon distribution function
($-0.8\leq y \leq 0.8$ for $\sqrt{s_{pp}}$ = 200 GeV with $f_g$ from  
Ref\cite{klm11}), is

\beq
\label{2dspplambda0}
     < \frac{d \sigma_{pp\rightarrow \Phi(\lambda=0)}}{dy}> &=& 
     A_\Phi \frac{1}{\bar{x}(y)} f_g(\bar{x}(y),2m)f_g(a/\bar{x}(y),2m) 
\frac{dx}{dy} \; ,
\eeq    
where, as is discussed above, $a= 4m^2/s$; with $m=1.5$  GeV for charmonium, 
and 5 GeV for bottomonium, and $A_\Phi=\frac{5 \pi^3 \alpha_s^2}{288 m^3 s}
<O_8^\Phi(^1S_0)>$ \cite{klm11}. For $\sqrt{s_{pp}}$ = 200 GeV 
$A_\Phi=7.9 \times 10^{-4}$nb for $\Phi$=$J/\Psi$ and $2.13 \times 10^{-5}$nb 
for $\Upsilon(1S)$; $a = 2.25 \times 10^{-4}$ for Charmonium and 
$2.5 \times 10^{-3}$ for Bottomium.

 The function $\bar{x}$, the effective parton x in a nucleus (A), is given in  
Refs\cite{vitov06,vitov09}:
\beq
\label{barx}
         \bar{x}(y)&=& x(y)(1+\frac{\xi_g^2(A^{1/3}-1)}{Q^2}) \nonumber \\
   x(y) &=& 0.5 \left[\frac{m}{\sqrt{s_{pp}}}(\exp{y}-\exp{(-y)})+
\sqrt{(\frac{m}{\sqrt{s_{pp}}}(\exp{y}-\exp{(-y)}))^2 +4a}\right] \;,
\eeq
with\cite{qiu04} $\xi_g^2=.12 GeV^2$. For $J/\Psi$  $Q^2=10 GeV^2$, so
$\bar{x}=1.058 x$ for Au and $\bar{x}=1.036 x$ for Cu, while for $\Upsilon(1S)$
$Q^2=100 GeV^2$, so $\bar{x}=1.006 x$ for Au and $\bar{x}=1.004 x$ for Cu.

\clearpage

From this we find the differential rapidity cross sections as shown Figs. 
34-41 for $J/\Psi, \Psi(2S)$ and 
$\Upsilon(1S),\Upsilon(2S), \Upsilon(3S)$ production via Cu-Cu and Au-Au 
collisions at RHIC (E=200 GeV), 
with $\Psi(2S),\Upsilon(3S)$ enhanced by $\pi^2/4$ as discussed above. The
absolute magnitudes are uncertain, and the shapes and relative magnitudes are 
our main prediction.
\vspace{2.8cm}

\begin{figure}[ht]
\begin{center}
\epsfig{file=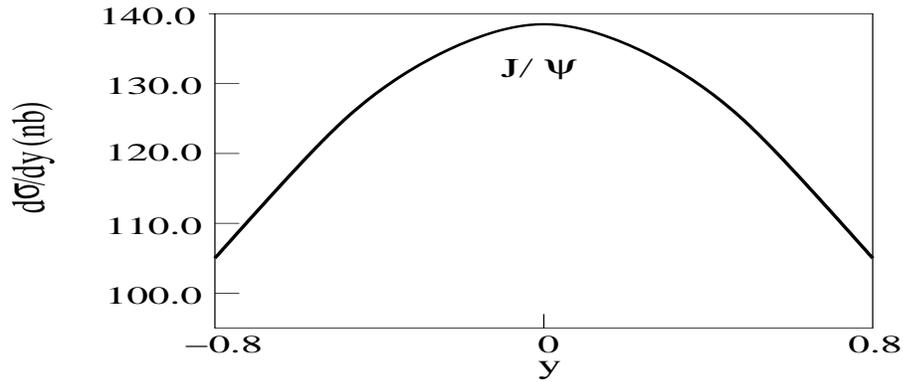,height=5 cm,width=12cm}
\caption{d$\sigma$/dy for 2m=3 GeV, E=200 GeV Cu-Cu collisions 
producing $J/\Psi$ with $\lambda=0$}
%\label{}
\end{center}
\end{figure}
\vspace{2.3cm}

\begin{figure}[ht]
\begin{center}
\epsfig{file=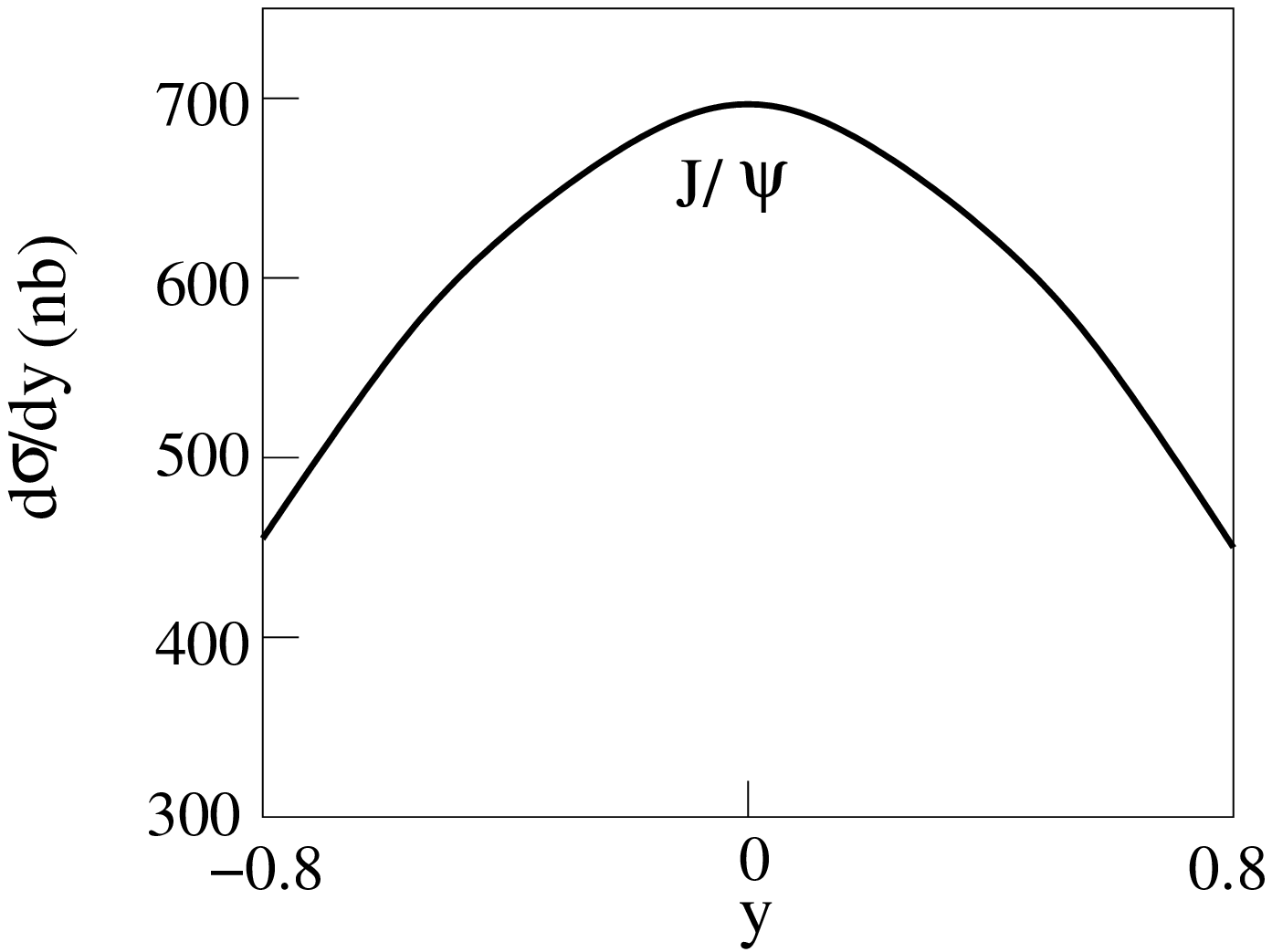,height=5 cm,width=12cm}
\caption{d$\sigma$/dy for 2m=3 GeV, E=200 GeV Au-Au collisions 
producing $J/\Psi$ with $\lambda=0$}
%\label{}
\end{center}
\end{figure}
\clearpage

\begin{figure}[ht]
\begin{center}
\epsfig{file=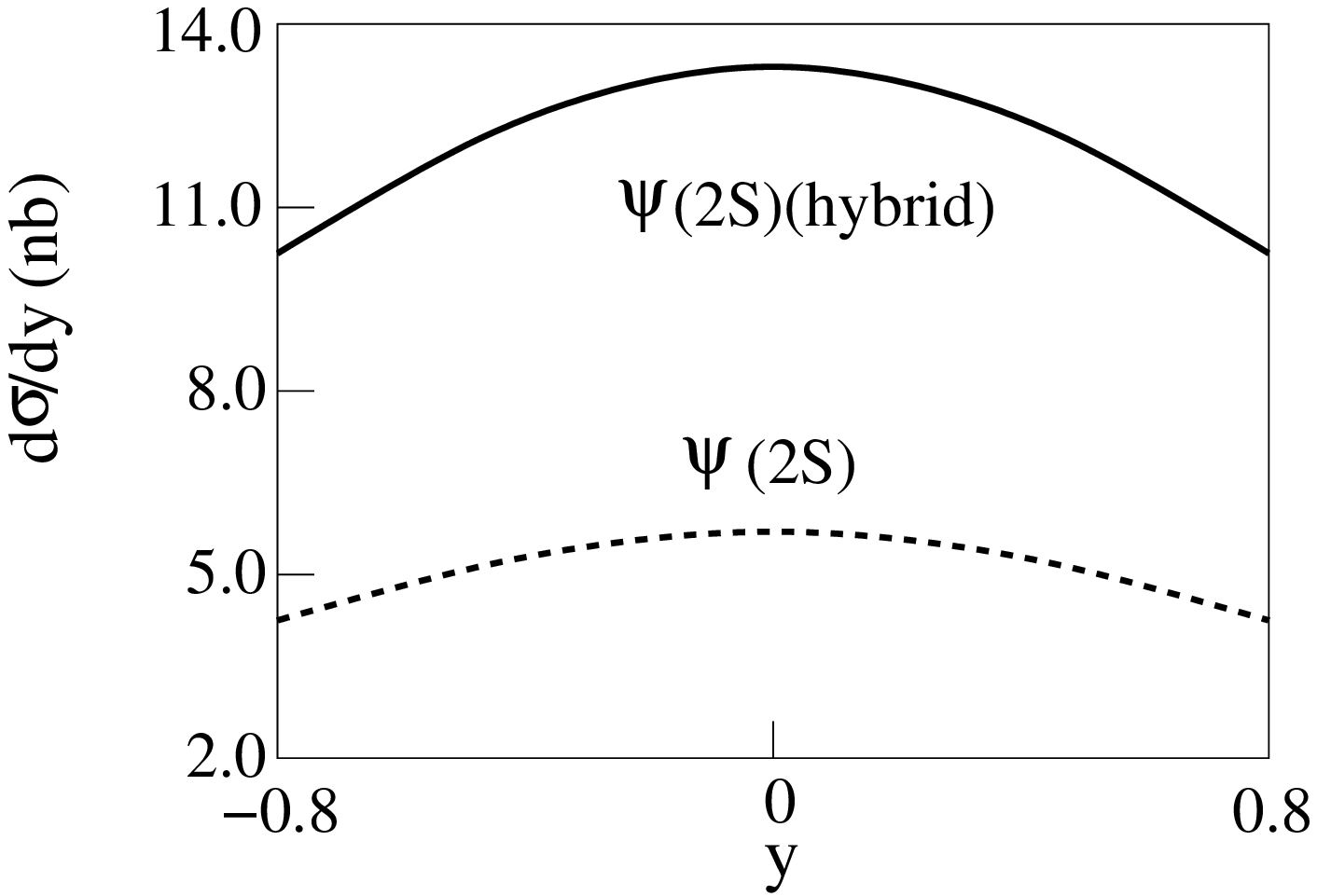,height=6 cm,width=12cm}
\caption{d$\sigma$/dy for 2m=3 GeV, E=200 GeV Cu-Cu collisions 
producing $\Psi(2S)$ with $\lambda=0$. The dashed curve is for the standard
$c\bar{c}$ model.}
%\label{}
\end{center}
\end{figure}

\vspace{4cm}

\begin{figure}[ht]
\begin{center}
\epsfig{file=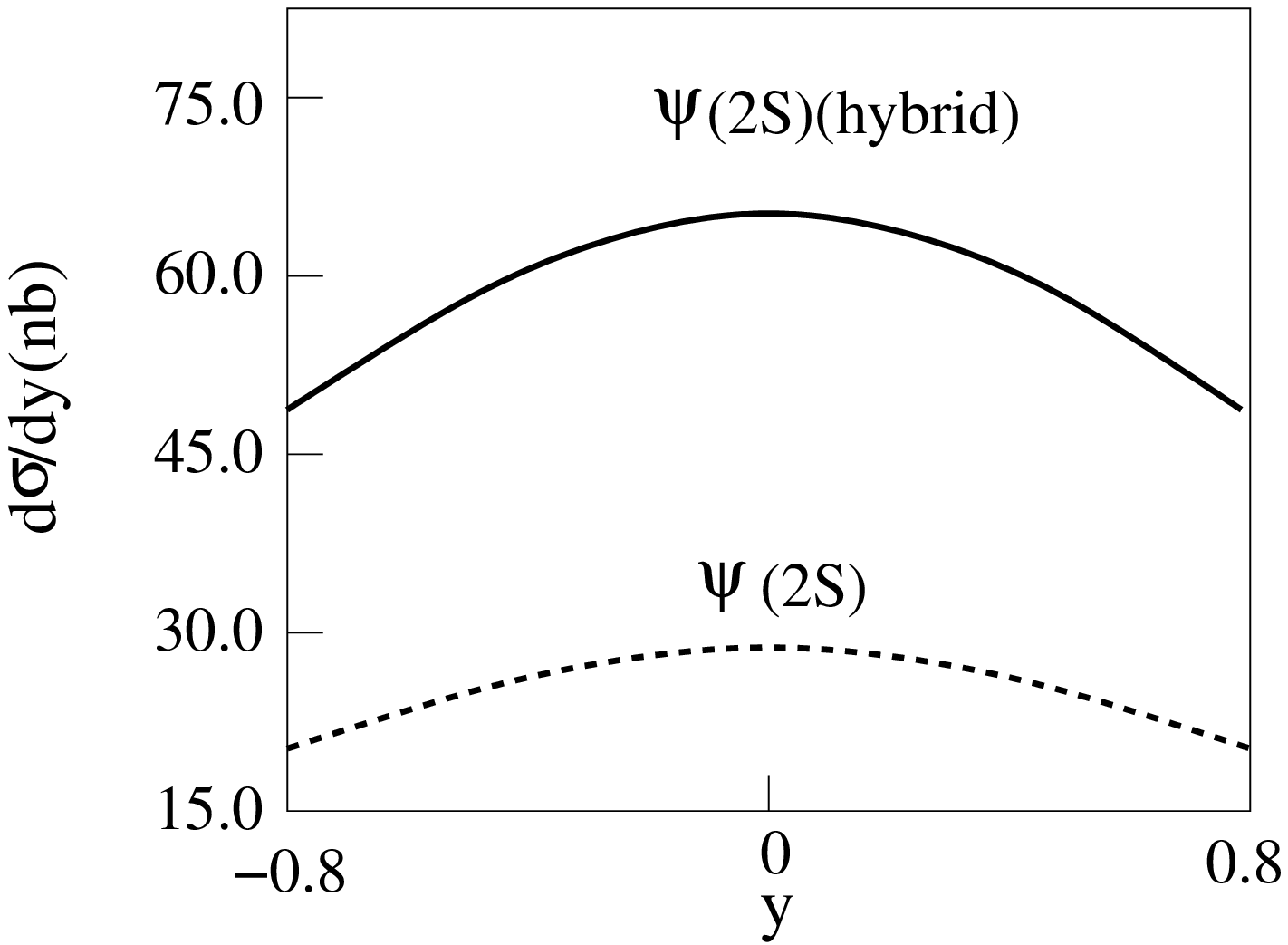,height=6 cm,width=12cm}
\caption{d$\sigma$/dy for 2m=3 GeV, E=200 GeV Au-Au collisions 
producing $\Psi(2S)$ with $\lambda=0$. The dashed curve is for the standard
$c\bar{c}$ model.}
%\label{}
\end{center}
\end{figure}
\clearpage

\begin{figure}[ht]
\begin{center}
\epsfig{file=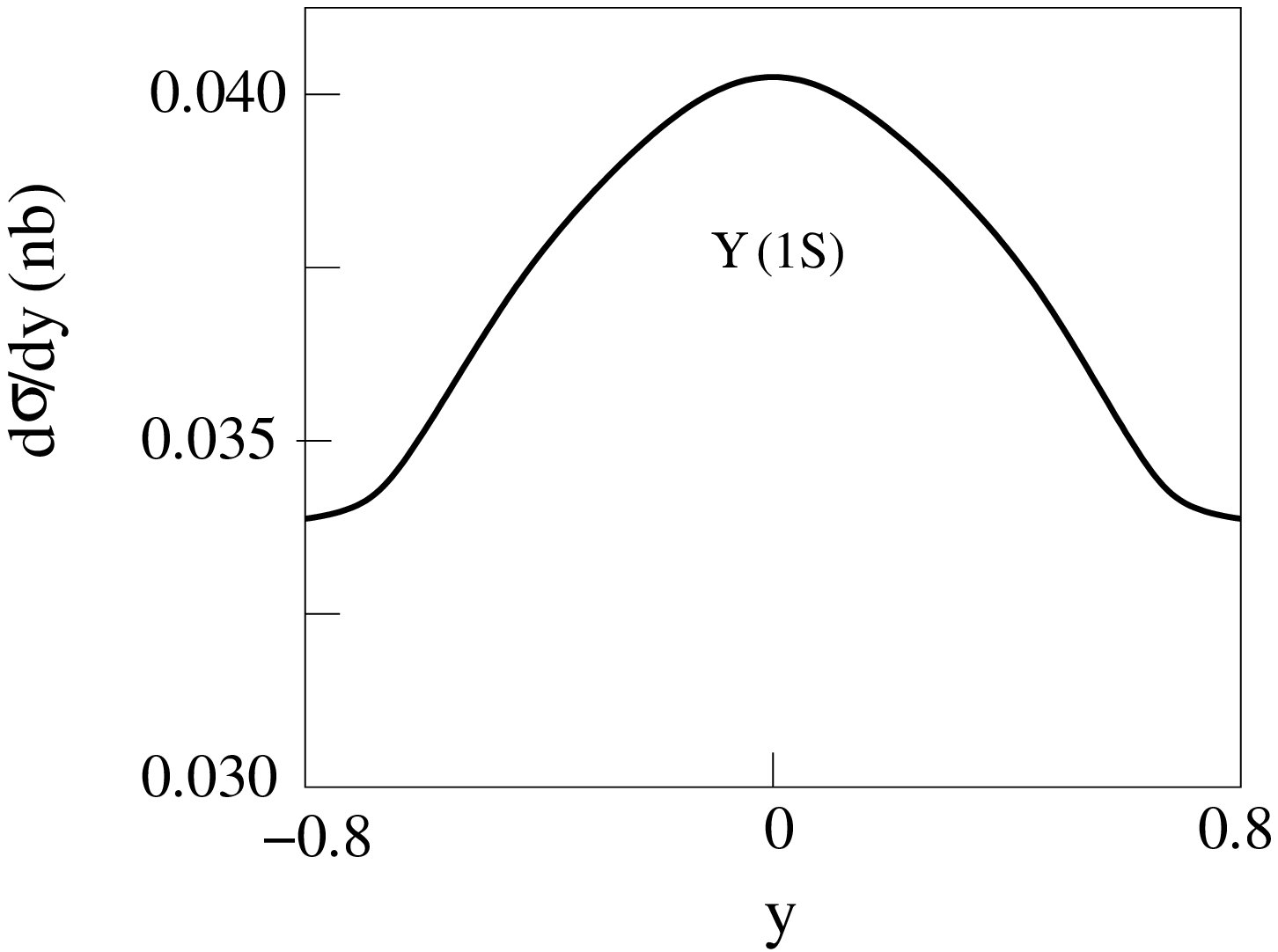,height=7 cm,width=12cm}
\caption{d$\sigma$/dy for 2m=10 GeV, E=200 GeV Cu-Cu collisions 
producing $\Upsilon(1S)$ with $\lambda=0$}
%\label{}
\end{center}
\end{figure}
\vspace{4 cm}

\begin{figure}[ht]
\begin{center}
\epsfig{file=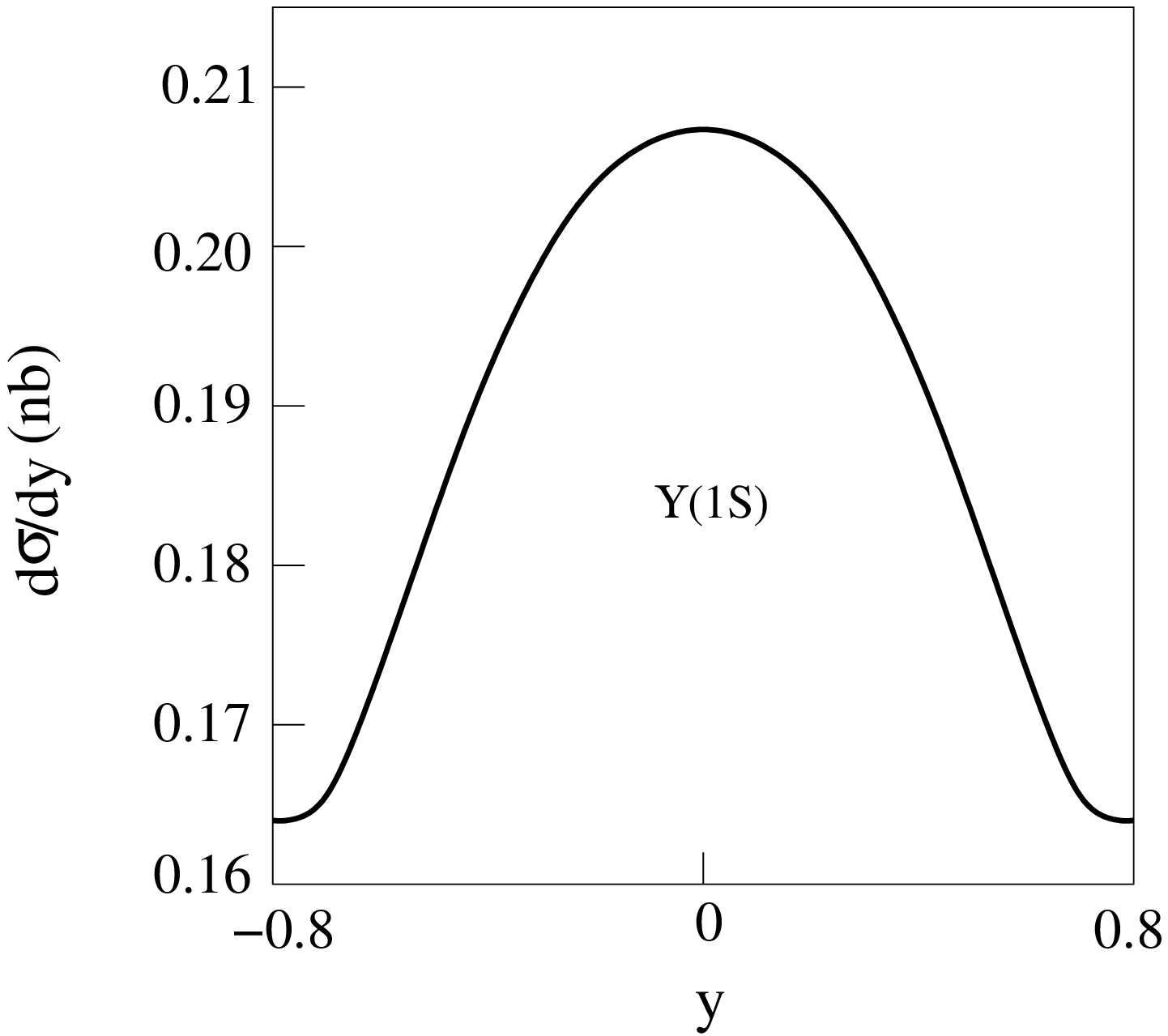,height=5 cm,width=12cm}
\caption{d$\sigma$/dy for 2m=10 GeV, E=200 GeV Au-Au collisions 
producing $\Upsilon(1S)$ with $\lambda=0$}
%\label{}
\end{center}
\end{figure}
\clearpage

\begin{figure}[ht]
\begin{center}
\epsfig{file=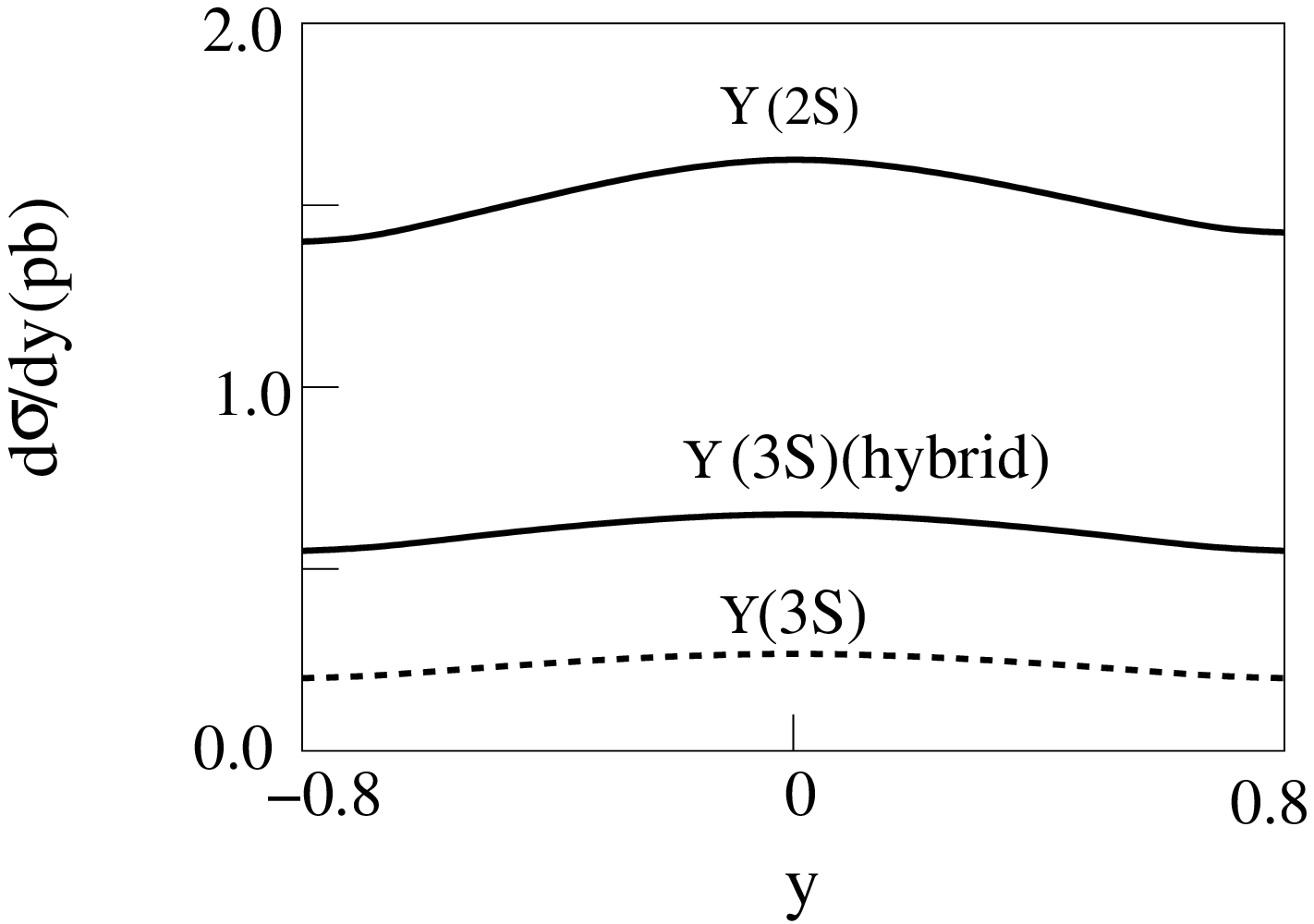,height=7 cm,width=12cm}
\caption{d$\sigma$/dy for 2m=10 GeV, E=200 GeV Cu-Cu collisions 
producing $\Upsilon(2S),\Upsilon(3S)$ with $\lambda=0$. For $\Upsilon(3S)$
the dashed curve is for the standard $b\bar{b}$ model.}
%\label{}
\end{center}
\end{figure}
\vspace{5 cm}

\begin{figure}[ht]
\begin{center}
\epsfig{file=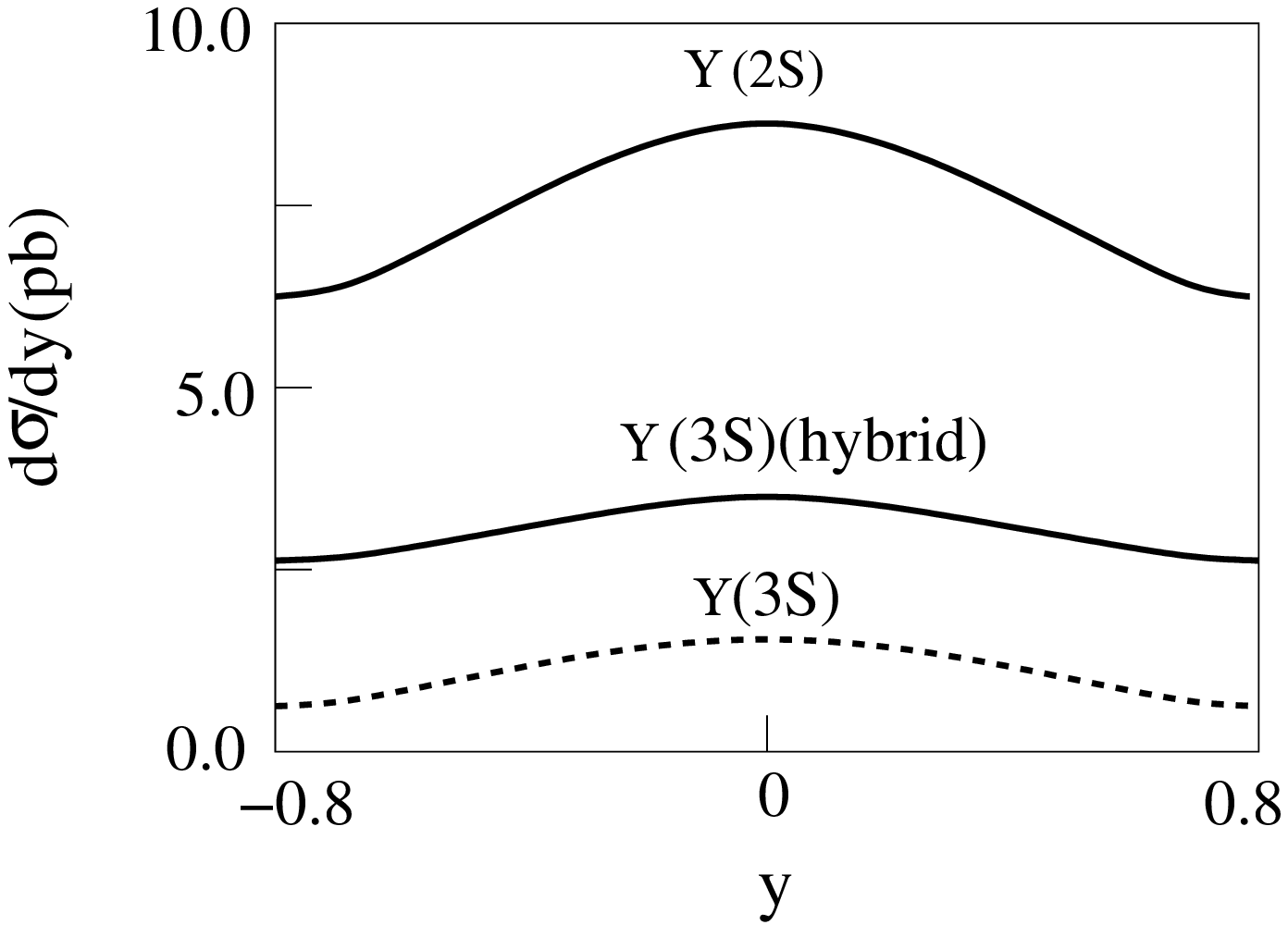,height=4 cm,width=12cm}
\caption{d$\sigma$/dy for 2m=10 GeV, E=200 GeV Au-Au collisions 
producing $\Upsilon(2S),\Upsilon(3S)$ with $\lambda=0$. For $\Upsilon(3S)$
the dashed curve is for the standard $b\bar{b}$ model.}
%\label{}
\end{center}
\end{figure}
\newpage

\subsection{Ratios of $\Psi'(2S)$ to $J/\Psi$ cross sections for A-A
collisions}

As discussed above,  for the standard (st), hybrid model(hy) one finds for 
p-p production of $\Psi'(2S)$ and $J/\Psi$
\beq
\label{ppratio}
    \sigma(\Psi'(2S))/\sigma(J/\Psi(1S))|_{st} &\simeq& 0.039 \nonumber \\
    \sigma(\Psi'(2S))/\sigma(J/\Psi(1S))|_{hy} &\simeq& 0.122 \; ,
\eeq
while the PHENIX experimental result for the ratio\cite{phoenix} $\simeq
0.18 \pm .04$. Therefore, the hybrid model is consistent with
experiment, while the standard model ratio is too small.

 The recent CMS/LHC result comparing Pb-Pb to p-p Upsilon
production\cite{cms11} found
\beq
\label{CMS2}
     [\frac{\Upsilon(2S) +\Upsilon(3S)}{\Upsilon(1S)}]_{Pb-Pb}/
    [\frac{\Upsilon(2S) +\Upsilon(3S)}{\Upsilon(1S)}]_{p-p} &\simeq& 
0.31^{+.19}_{-.15} \pm .013(syst) \; ,
\eeq
while in the  work discussed previously on $p-p$ collisions the ratio 
$\sigma(\Upsilon(3S))/\sigma(\Upsilon(1S))|_{p-p}$
of the standard $|b\bar{b}>$ model was $4/\pi^2 \simeq 0.4$ of the hybrid
model. This suggests a suppression factor for $\sigma(b\bar{b}(3S))/
\sigma(b\bar{b}(1S))$, or $\sigma(c\bar{c}(2S)/\sigma(c\bar{c}(1S))$ of 
0.31/.4 as these components travel 
through the QGP; or an additional factor of 0.78 for $\Psi'(2S)$ to $J/\Psi$
production for $A-A$ vs $p-p$ collisions. Therefore from Eq(\ref{ppratio}) 
one obtains the estimate using the mixed hybrid theory for this ratio
\beq
  \sigma(\Psi'(2S))/\sigma(J/\Psi(1S))|_{A-A{\rm \; collisions}} &\simeq& 
0.10 
\eeq
\subsection{Ratios of $\Upsilon(2S)$ and $\Upsilon(3S)$
 to $\Upsilon(1S)$ cross sections for Pb-Pb vs p-p collisions}

 As pointed out in Eq(\ref{ppUpratio}), $\sigma(\Upsilon(2S))/
\sigma(\Upsilon(1S))|_{standard} \simeq
\sigma(\Upsilon(2S))/\sigma(\Upsilon(1S))|_{hybrid} \simeq 0.039$,
$\sigma(\Upsilon(3S))/\sigma(\Upsilon(1S))|_{standard} \simeq .0064$,
Although the ratio $\sigma(\Upsilon(3S))/\sigma(\Upsilon(1S))$ is
difficult to measure, as pointed out above, the ratios of cross sections
for $\sigma(\Upsilon(2S))/\sigma(\Upsilon(1S))$ and 
$\sigma(\Upsilon(3S))/\sigma(\Upsilon(1S))$ for A-A vs p-p can be measured.

  The recent CMS experiment's main objective\cite{cms12} is to test for 
$\Upsilon$ suppression in PbPb collisions, with estimates of the following
quantities:
\beq
\label{cmsz12}
  && \frac{[\Upsilon(2S)/\Upsilon(1S)]_{PbPb}}
 {[\Upsilon(2S)/\Upsilon(1S)]_{pp}} \nonumber \\
  &&\frac{[\Upsilon(3S)/\Upsilon(1S)]_{PbPb}}
 {[\Upsilon(3S)/\Upsilon(1S)]_{pp}} \; .
\eeq
  The studies of A-A collisions for Bottomonium states, which cannot be
carried out at RHIC but are an important part of the LHC CMS program,
is expected to be carried out in future research. 
\newpage
\subsection{Creation of the QGP via A-A collisions}

A main goal of the study of heavy quark state production in A-A colisions
is the detection of the Quark Gluon Plasma. The energy of the atomic nuclei
must be large enough so just after the nuclei collide the temperature is
that of the unverse about $10^{-5}$ seconds after the Big Bang, when the
universe was too hot for protons or neutrons and consisted of quarks
and gluons (the constituents of proton and nucleons)-the QGP. As Figure 42
illustrates, the emission of mixed hybrid mesons, the $\Psi(2S)$
and $\Upsilon(3S)$ as discussed above, with active gluons, could be a
signal of the formation of the QGP.

\begin{figure}[ht]
\begin{center}
\epsfig{file=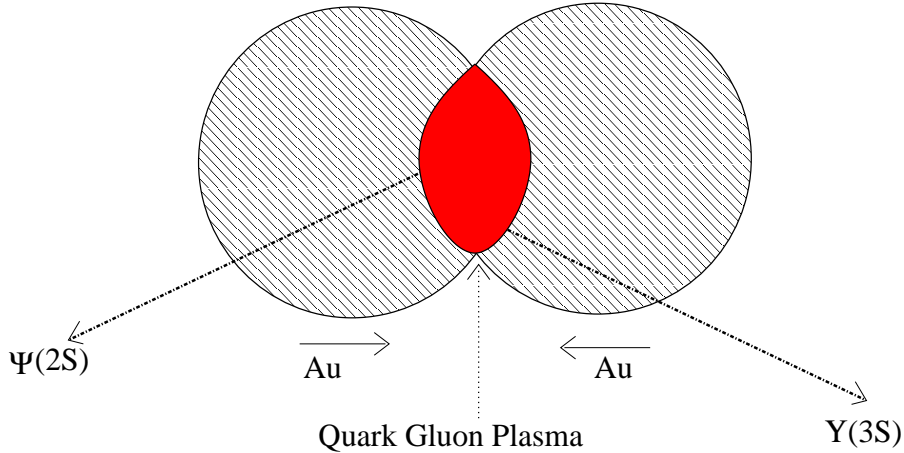,height=6 cm,width=12cm}
\caption{ Au-Au collisions producing $\Psi(2S)$ and $\Upsilon(3S)$ from
theQGP.}
%\label{}
\end{center}
\end{figure}

\subsection{Conclusions for Heavy-quark state production in A-A 
collisions at $\sqrt{s_{pp}}$=200 GeV}
  
The differential rapidity cross sections for
$J/\Psi, \Psi(2S)$ and $\Upsilon(nS)(n=1,2,3)$ production via Cu-Cu and
Au-Au collisions at RHIC (E=200 GeV) were calculated using $R_{AA}$, 
the nuclear modification factor, $N^{AA}_{bin}$ the binary collision number,
and the gluon distribution functions. This should give some guidance for 
future RHIC experiments, although at the present time the 
$\Upsilon(nS)$ states cannot be resolved.

The ratio of the production of $\sigma(\Psi'(2S))$, which in the mixed
hybrid theory is 50\% $c\bar{c}(2S)$ and 50\% $c\bar{c}g(2S)$ with a 
$10\%$ uncertainty, to
$J/\Psi(1S)$, which is the standard $c\bar{c}(1S)$, could be an important
test of the production of the Quark-Gluon Plasma. Using the hybrid model
and suppression factors from previous theoretical estimates and experiments
on $\Upsilon(mS)$ state production at the LHC, the ratio 
of $\Psi'(2S)$ to $J/\Psi(1S)$ production at RHIC via A-A collisions is
estimated to be about $0.52\pm 0.05$. In future studies at BNL and the 
LHC-CERN the study of RHIC producing $\Psi'(2S)$ and $\Upsilon(3S)$ mixed
hybrid meson could be a method for determining the creation of the QGP.

\subsection{$J/\Psi$ state production in Pb-Pb collisions
 at $\sqrt{s_{NN}}$=2.76 TeV}

 There have also been a number of experiments by the ALICE Collaboration
on the production of $J/\Psi$ via Pb-Pb collisions at 2.76 TeV\cite{ALICE12,
das12,ALICE14} which has measured $R_{AA}$ and other aspects of A-A
collisions needed to establish the detection of the QGP. Since the present
review is mainly focused on experimental tests of the mixed hybrid theory,
with present detectors and also future LHC upgrades,
we do not discuss these experimental publications in further detail.
 
\section{Production of Charmonium and Upsilon States via
Fragmentation}

 In the previous sections we reviewed the production of $c\bar{c}$ and
$b\bar{b}$ states via p-p and A-A collisions. In this section we review
the production of $|c\bar{q}>$ and  $|b\bar{q}>$, with $q$ a light quark. 
Therefore the dominant octet processes illustrated in Figure 12, which produce 
$Q\bar{Q}$ states
are not sufficient. To produce a $Q\bar{q}$ state, with $Q=c{\rm \;or\;}b$
and $q$ a light quark, one needs the quark fragmentation processes, which
was introduced for the study of $Z^0$ (a weak gauge boson) decay\cite{bcy93}. 
This is illustrated in Figure 43.

\begin{figure}[ht]
\begin{center}
\epsfig{file=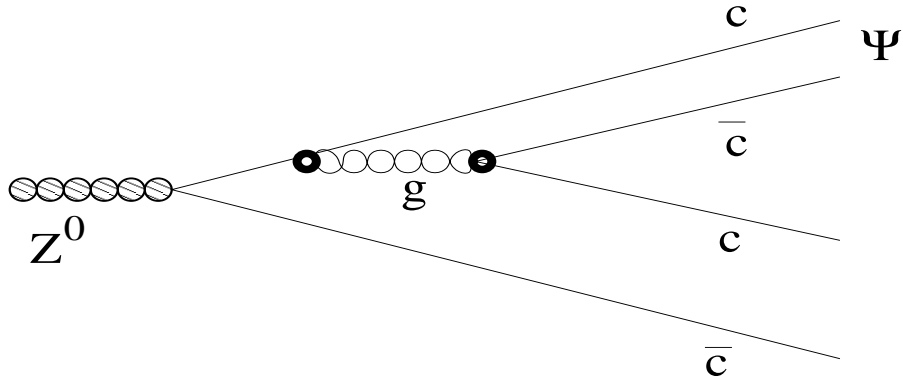,height=5 cm,width=12cm}
\caption{ Quark fragmentation for $Z^0 \rightarrow \Psi+ c\bar{c}$}
%\label{}
\end{center}
\end{figure}

  The fragmentation probability which is used in the production of D-mesons
via p-p collisions discussed in the following subsection was calculated
by Bratten $et. al.$\cite{bcfy95}
\vspace{3mm}

  Gluon fragmentation into heavy quarkonium calculated in Ref\cite{by93} is
illustrated in Figure 44. Although it is important for some charmonium or
bottonium state production, we do not use it in the present review.
\newpage

\begin{figure}[ht]
\begin{center}
\epsfig{file=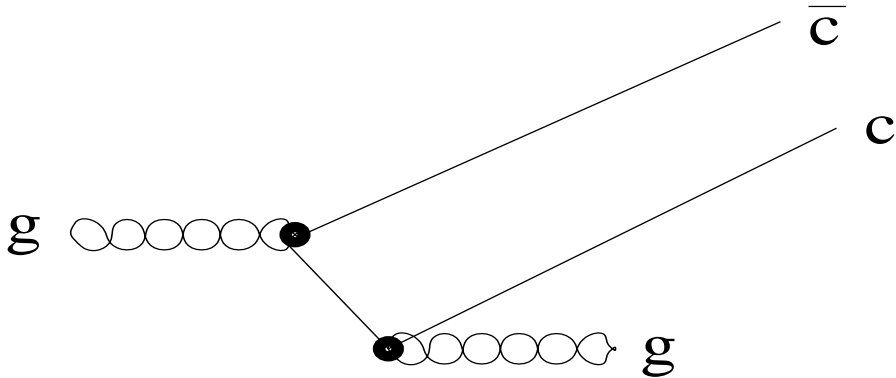,height=5 cm,width=12cm}
\caption{Gluon fragmentation into $c\bar{c}g$}
%\label{}
\end{center}
\end{figure}

\subsection{D Production In p-p and d-Au Collisions }

 In this subsection the production of $D^+(c\bar{d}),D^o(c\bar{u})$ Charm
mesons via unpolarized p-p and d-Au collisions at 200 GeV, based on recent
research that has not yet been published\cite{klm15}, is discussed. 
The main new aspect of the present work is that while a gluon can produce a
$c\bar{c}$ or $b\bar{b}$ state, it cannot directly produce a $c\bar{d}$.
A fragmentation process converts a $c\bar{c}$ into a $c\bar{d}-d\bar{c}$,
for example. We use the fragmentation probability, 
$D_{c \rightarrow c\bar{q}}$ of Bratten et. al.\cite{bcfy95}, illustrated in 
Figure 45.

\begin{figure}[ht]
\begin{center} 
\epsfig{file=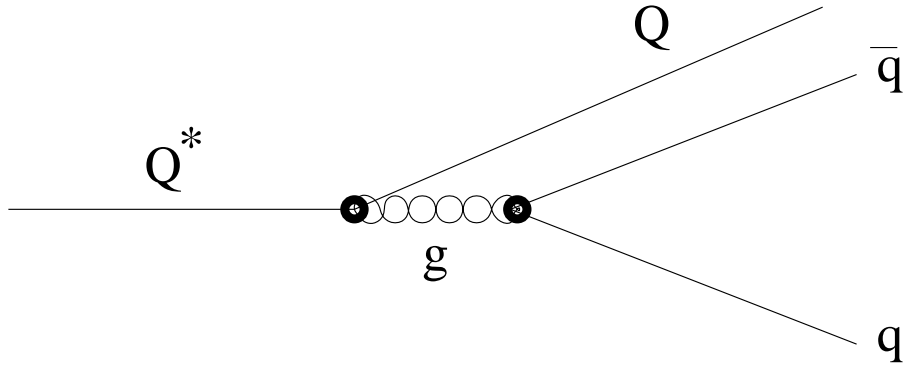,height=5 cm,width=12cm}
\caption{ Quark fragmentation for $Q^* \rightarrow (Q\bar{q})+q)$}
%\label{}
\end{center}
\end{figure}
\subsubsection{Differential $pp\rightarrow DX$ cross section} 

Using what in Ref\cite{ns06} is called scenerio 2, the production
cross section with gluon dominance for DX is
\beq
\label{sppDz}
  \sigma_{pp\rightarrow Dx} &=&  \int_a^1 \frac{d x}{x} 
f_g(x,2m)f_g(a/x,2m) \sigma_{gg\rightarrow DX}\; ,
\eeq
with\cite{bcfy95}
\beq
\label{sggDX}
 \sigma_{gg\rightarrow DX}&=& 2 \sigma_{gg\rightarrow c\bar{c}}
D_{c\rightarrow c\bar{q}} \; ,
\eeq
where $\sigma_{gg\rightarrow c\bar{c}}$ is similar to the charmonium production
cross section in Ref\cite{klm11} and $D_{c\rightarrow c\bar{q}}$ is the total
fragmentation probability.

For $E=\sqrt{s}$=200 GeV the gluon distribution funtion is
\beq
\label{fg}
 f_g(y) &=& 1334.21 - 67056.5 x(y) + 887962.0 x(y)^2
\eeq
From Ref\cite{bcfy95}, using the light quark mass=(up quark mass+down 
quark mass)/2=3.5 Mev.
\beq
\label{Dccq}
    D_{c \rightarrow c\bar{q}}&=& 9.21 \times 10^{5} \alpha_s^2 |R(0)|^2/\pi 
 \; ,
\eeq
in units of $(1/GeV^3)$, with $\alpha_s=.26$. For a 1S state
$|R(0)| ^2 = 4/(a_o)^3$. For a $c\bar{q}$ state, $(1/a_o)=m_q\simeq 3.5$ MeV. 
Therefore,
\beq
\label{R(0)}
   |R(0)| ^2&\simeq& 1.71 \times 10^{-7} {\rm \;\;(GeV)^3} \nonumber \\
    D_{c \rightarrow c\bar{q}}&\simeq& 3.39 \times 10^{-3} \; .
\eeq

The calculation is similar to that in Ref\cite{klm11}.
\beq
\label{ppDX}
  \frac{d\sigma_{pp\rightarrow DX}}{dy}&=& Acc*f_g(x(y),2m) f_g(a/x(y),2m) 
\frac{dx(y)}{dy} \frac{1}{x(y)} D_{c \rightarrow c\bar{q}} \; ,
\eeq
with $y$ and $x(y)$ defined above and
$Acc$ is the matrix element for charmonium production\cite{klm11}
with an effective mass $ms$=1.5 GeV
\beq
\label{Acc}
       Acc&=& 7.9*10^{-4} nb \; .
\eeq

  From Eqs(\ref{ppDX},\ref{Acc}) one finds $\frac{d\sigma_{pp\rightarrow DX}}{dy}$ 
shown in Figure 44 after the next subsubsection.

\subsubsection{Differential $dAu\rightarrow DX$ cross section}

In this subsubsection we estimate the production of $D^+, D^0$ from d-Au 
collisions, using the methods given in Ref.\cite{klm14} for the estimate of 
production of $\Psi$ and $\Upsilon$ states via Cu-Cu and Au-Au collisions 
based on p-p collisions.

  The differential rapidity cross section for D+X production via d-Au
collisions is given by $ \frac{d\sigma_{pp\rightarrow DX}}{dy}$ with modification
described in Ref.\cite{klm14} for Cu-Cu and Au-Au collisions:
\beq
\label{sigmadAu}
  \frac{d\sigma_{dAu\rightarrow DX}}{dy}&=& R_{dAu}N^{dAu}_{coll}
\left (\frac{d\sigma_{pp\rightarrow DX}}{dy} \right)  \; ,
\eeq
where $R_{dAu}$ is the nuclear-modification factor, $N^{dAu}_{coll}$ is the
number of binary collisions, and $\left (\frac{d\sigma_{pp\rightarrow DX}}{dy} 
\right)$ is the differential rapidity cross section for $DX$ production via
nucleon-nucleon collisions in the nuclear medium.

  $\left (\frac{d\sigma_{pp\rightarrow DX}}{dy} \right)$ is given by Eq(\ref{ppDX})
with $x(y)$ replaced by the function $\bar{x}$  (see (Eq(\ref{barx})), the 
effective parton x in the nucleus Au.
\newpage

  In Ref.\cite{phenix14} the  quantities  $R_{dAu}$ and $N^{dAu}_{coll}$
(called $R_{dA}$ and $<N_{coll}>$ in that article) were estimated from
experiments on p+p and d+Au collisions. From that reference
$R_{dAu} \simeq 1.0$ and  $N^{dAu}_{coll}\simeq 10.0$.

  From Eqs(\ref{ppDX},\ref{sigmadAu}), one obtains the differential 
rapidity cross section for D+X production via dAu collisions. In 
Figure 46  $\frac{d\sigma_{pp\rightarrow DX}}{dy}$ and
 $\frac{d\sigma_{dAu\rightarrow DX}}{dy}$ are shown
\vspace{5cm}

\begin{figure}[ht]\begin{center}
\epsfig{file=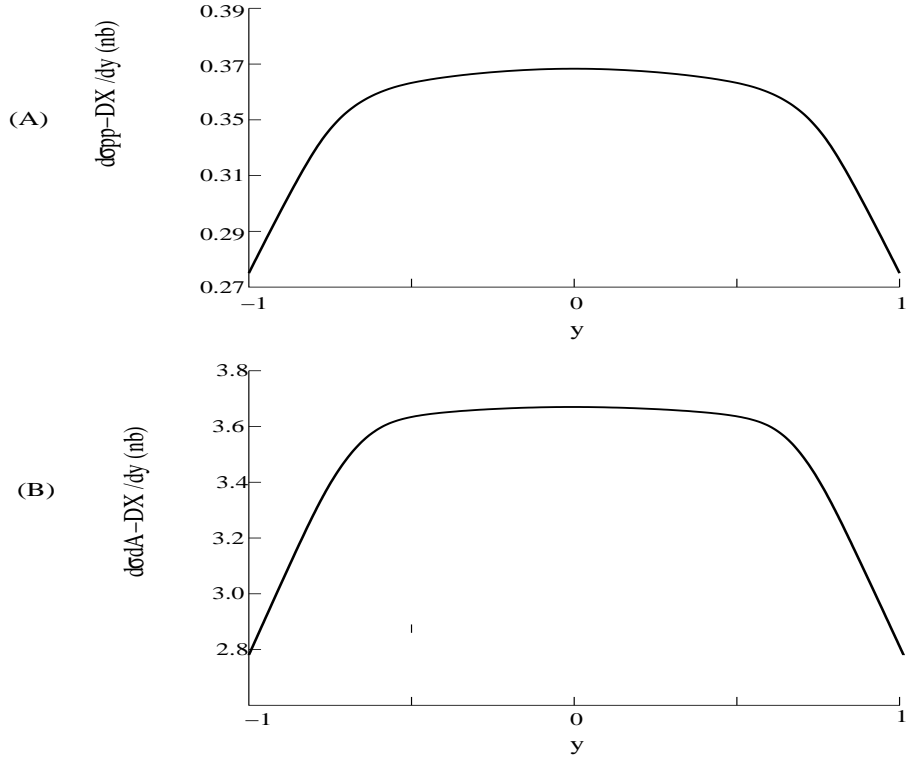,height=10cm,width=12cm}
\end{center}
\caption{$d\sigma/dy$ with E=200 GeV for (A) unpolarized p-p collisions and 
(B) d-Au collisions producing D+X} 
\end{figure}

\newpage
  A number of experiments have measured $\sigma_{c\bar{c}}$ cross 
sections at $\sqrt{s_{pp}}$=200 GeV\cite{phenix06,phenix-2-07,star07,phenix09}.
Theoretical estimates of heavy quark state production via p-p collisions
at RHIC and LHC energies were made almost two decades ago\cite{ggm95}.
More recently estimates of $D$ production were made from data on d-Au
collisions at $\sqrt{s_{NN}}$= 200 GeV\cite{star05}. Experimental measurements
of $D^+, D^-, D^0$ production via p-p and d-Au collisions are expected in the 
future.  

\section{Sivers and Collins Fragmentation Functions}

  The E1039 Collaboration, see Ref\cite{LOI13-15} for the Letter of Intent, 
plans to carry out a Drell-Yan experiment with a polarized Proton target, with
the main objective to measure the Sivers function\cite{siv90}. This Letter of 
Intent has motivated this brief review of Sivers and Collins symetries and 
fragmentation functions.

 A number of Deep Inelastic Scattering experiments\cite{hermes09,compass09,
jlab11} have measured non-zero values for the Sivers Function. See these 
references for references  to\ earlier experiments.  Another important 
function is the Collins fragmentation function\cite{col93}, which describes 
the fragmentation of a transversly polarized quark into an unpolarized hadron, 
such as a pion. The Sivers and Collins functions are defined by the target 
assymmetry, $A(\phi,\phi_S$), in the scattering of an unpolarized lepton beam 
by a transversely polarized target\cite{jlab11}:
\beq
\label{sivcol}
     A(\phi,\phi_S)&\simeq&A_C(\phi,\phi_S)sin(\phi+\phi_S)+
A_S(\phi,\phi_S)sin(\phi-\phi_S) \;,
\eeq
where $A_C,A_S$ are the Collins,Sivers functions with $\phi$ the azimuthal 
angle and $\phi_S$ the aximuthal angle with respect to the lepton beam.

\subsection{ Sivers Function}

  The Sivers term of the cross section for the production of hadrons using an 
unpolarized lepton beam on a transversely polarized target is\cite{hermes09}
\beq
\label{sivers}
 \sigma_{S}(\phi,\phi_S)&=& \sigma_{UU} S_T(x)[2n(<sin(\phi-\phi_S)>_{UT}\times
sin(\phi-\phi_S) +...] \;,
\eeq
where $\phi$ and $\phi_S$ were defined above. $\sigma_{UU}$ is the $\phi$-
independent part of the polarization-independent cross section; and $UT$ denotes
the unpolarized beam with transverse target polarization w.r.t. the virtual
photon direction. $S_T(x)$ is the Sivers 
Function with $x=-q^2/(2P\cdot q)$ and $P$ is the four-momentum of the target 
proton. As mentioned above, a number of Deep Inelastic 
Scattering experiments have measured $S_T(x)$ and obtained non-zero values.
See, e.g., Ref\cite{hermes09} for a discussion of the Sivers Function 
in terms of experimental cross sections.

\subsection{Collins Fragmentation Function}

The definition of the Collins Function is similar to that of the Sivers
Function in Eq(\ref{sivers}\cite{kang15}:
\beq
\label{collins}
 \sigma_C(\phi,\phi_S)&\propto& F_{UU}(1+A_{UT}\times sin(\phi+\phi_S))
 \;,
\eeq
with $\phi$ and $\phi_S$ defined above, $F_{UU}$ the spin-averaged structure 
function, and $A_{UT}$ the asymmetry that can be 
calculated from quark distribution and fragmentation, which is now discussed 
briefly .

From Ref\cite{bggm08} the fragmentation function to produce a hadron, $h$ from
a transversely polaried quarek in $e^+ e^-$ annihilation is
\beq
\label{polqfrag}
 D_{h q^{\uparrow}}(z,k_T^2)&=& D^q(z,k_T^2)+H^q(z,k_T^2)\frac{k\times k_T/k)
\cdot s_q}{z M_h} \; ,
\eeq
where $M_h$ is the hadron mass, k is the quark momentum, $s_q$ is the quark
spin vector, $k_T$ the hadron momentum transverse to k, and z the light-cone
momentum fraction of h wrt the fragmenting quark. The Collins fragmentation
function is $H^q$.

  The calcultion of $D_{h q^{\uparrow}}$ is similar to the fragmentation
function for D production in p-p collisions\cite{klm15} discussed above.
An estimate of the Collins Fragmentation Function is being carried
out\cite{lsk15}.

\section{Brief Overview}

The theoretical basis for production of heavy quark states via p-p collisions,
using the standard model for $J/\Psi,\Upsilon(1S),\Upsilon(2S)$ states
and a mixed hybrid theory for  $\Psi(2S),\Upsilon(3S)$ using QCD and QCD
Sum Rules has been established by comparison with many experiments. For
detection of the Quark-Gluon plasma, a main objective of RHIC and an
important objective for the LHC, production of heavy quark states via A-A
collisions is required. This is much more complicated, but there has
been a great deal of progress in both experiment and theory. The detection of 
the Quark-Gluon Plasma via A-A collisions is closer to realization with this 
improved theory.

Also, the theory of production of open charm and bottom meson via p-p and A-A 
collisions is now greatly improved using the theory of Fragmentation. Deep 
inelastic experiments for measuring the Sivers and Collins Fragmentation
functionns are being carried out and are planned for the future.
  
\Large{{\bf Acknowledgements}}

\vspace{5mm}
\normalsize 
Author D.D. acknowledges the facilities of Saha Institute of Nuclear Physics, 
Kolkata, India. Author L.S.K. acknowledges support from the P25 group at Los 
Alamos National laboratory.

\end{document}